\documentclass[aps,prx,twocolumn,superscriptaddress,showpacs,amsmath,amssymb,longbibliography,nofootinbib]{revtex4-1}
\usepackage[english]{babel}
\usepackage{amsmath}
\usepackage{bm}
\usepackage{graphicx,bbm} 
\usepackage{times}
\usepackage{epsfig} 
\usepackage{soul}
\usepackage[utf8]{inputenc}
\usepackage[colorlinks,linkcolor=blue,citecolor=blue,urlcolor=blue]{hyperref}
\usepackage{color}
\usepackage{latexsym}
\usepackage{ulem}
\definecolor{nred} {RGB}{224,0,0}
\definecolor{nblue} {RGB}{28,130,185}
\definecolor{dgreen} {RGB}{78,138,21}
\definecolor{norange}{RGB}{230,120,20}
\definecolor{mypurple}{RGB}{100,0,200}

\usepackage{bbding}
\usepackage{pifont}
\usepackage{wrapfig}
\usepackage{amsmath}
\usepackage{verbatim}
\usepackage{amssymb}
\usepackage{inputenc}
\usepackage{textcomp}
\usepackage{appendix}
\usepackage{mathrsfs}
\usepackage{float}
\usepackage[margin=10pt,font=small,labelfont=bf, labelsep=endash,format=plain, justification=RaggedRight]{caption} [2022/02/20]
\usepackage{xfrac}
\restylefloat{table}
\usepackage{subfig}
\usepackage{mathtools}
\usepackage{tabularx}
\usepackage{bigints}
\usepackage{wasysym}

\newcommand{\be}[1]{\begin{equation}\label{#1} }
\newcommand{\ee}{\end{equation}}
\newcommand{\bea}[1]{\begin{eqnarray}\label{#1} }
\newcommand{\eea}{\end{eqnarray}}

\renewcommand{\a}{\alpha}

\renewcommand{\b}{\beta}

\newcommand{\s}{\sigma}

\newcommand{\bes}{\begin{subequations}}
\newcommand{\ees}{\end{subequations}}

\newcommand{\cH}{\mathcal{H}}

\def\ket#1{\left|#1\right\rangle}
\definecolor{nblue} {RGB}{28,130,185}
\definecolor{dgreen} {RGB}{78,138,21}
\definecolor{norange}{RGB}{230,120,20}
\definecolor{mypurple}{RGB}{100,0,200}
\usepackage[margin=0.75in]{geometry}
\hypersetup{colorlinks=false, linkcolor=blue, citecolor=red}
\begin{document} 
\title{Quantum state complexity meets many-body scars}
\author{Sourav Nandy}
\email{sourav.nandy@ijs.si}
\affiliation{Jo\v zef Stefan Institute, SI-1000 Ljubljana, Slovenia}
\author{Bhaskar Mukherjee}
\email{b.mukherjee@ucl.ac.uk}
\affiliation{Department of Physics and Astronomy, University College London, Gower Street, London WC1E 6BT, United Kingdom}
\author{Arpan Bhattacharyya}
\email{abhattacharyya@iitgn.ac.in}
\affiliation{Indian Institute of Technology Gandhinagar, Gujarat-382355, India}
\author{Aritra Banerjee}
\email{aritra.banerjee@oist.jp}
\affiliation{Okinawa Institute of Science and Technology, \\1919-1 Tancha, Onna-son, Okinawa 904-0495, Japan }

\date{\today}
\begin{abstract}
Scar eigenstates in a many-body system refers to a small subset of non-thermal finite energy density eigenstates embedded into an otherwise thermal spectrum. This novel non-thermal behaviour has been seen in recent experiments simulating a one-dimensional PXP model with a kinetically-constrained local Hilbert space realized by a chain of Rydberg atoms. We probe these small sets of special eigenstates starting from particular initial states by computing the spread complexity associated to time evolution of the PXP hamiltonian. Since the scar subspace in this model is embedded only loosely, the scar states form a weakly broken representation of the Lie Algebra. We demonstrate why a careful usage of the Forward Scattering Approximation (or similar strategies thereof) is required to extract an appropriate set of Lanczos coefficients in this case as the consequence of this approximate symmetry. This leads to a well defined notion of a closed Krylov subspace and consequently, that of spread complexity. We show how the spread complexity shows approximate revivals starting from both $|\mathbb{Z}_2\rangle$ and $|\mathbb{Z}_3\rangle$ states and how these revivals can be made more accurate by adding optimal perturbations to the bare Hamiltonian. We also investigate the case of the vacuum as the initial state, where revivals can be stabilized using an iterative process of adding few-body terms.  

\end{abstract}

\maketitle

%------------------------------------------------------------------------------------
\section{Introduction}
%\section{Quantum Scars}
The remarkable phenomenon of quantum scarring was first shown to occur in the case of quantum counterpart of classically chaotic billiard systems \cite{HellerPRL}. One can think of scar states as eigenstates of a system with a potential that show anomalously high probability density around the classical periodic trajectories associated to the said potential. Thus one would say, the classical system has left some memories, or has `scarred' the analogous quantum system. This is intriguing in the sense that it imparts a notion of regularity in a chaotic system, and gives rise to very special `protected' states, often referred to as the Quantum Single Particle Scars \cite{KAPLAN1998171} in the literature. The many-body cousins of these systems have generated newfound interest recently with the observations in a Rydberg chain experiment \cite{Bernien2017} that mimics the scarring phenomena. This has led to intense efforts to explore the physics of many-body scarred states \cite{scar1,Turner2018, Choi2019PRL, Lin2019, SenPRL2021,Khemani2020, MukherjeePRB2019, Klebanov2020PRL}\footnote{This list is in no way exhaustive, the reader is directed to the reviews \cite{MoudyagalyaRev,roderichrev} for a more complete list of references.}.
\medskip

Another important aspect of such states is their connection to ergodicity breaking phenomena, beyond the known paradigms of integrable models with inifinitely many conserved charges \cite{Caux:2010by} and many-body localization (MBL) \cite{MBL1,MBL2,MBL3,MBL4}. In the context of ergodicity of many-body systems, Eigenstate Thermalization Hypothesis (ETH) \cite{deutsch1,Srednicki:1994mfb,Rigol1}posits that all the bulk eigenstates of a generic many-body quantum system are thermal and this leads to eventual thermalization of a generic many-body state in non-equilibrium dynamics  \cite{Calabrese:2005in}. 
%Eigenstate Thermalization Hyptothesis (ETH) \cite{deutsch1,Srednicki:1994mfb,Rigol1} dictates that  expectation values of  quasi-local observables in an isolated interacting generic (i.e. systems which have no other conserved quantity except energy) many-body system looks locally thermal. Or equivalently speaking, all ergodic quantum systems by defintion thermalize at the late times in the thermodynamic limit \cite{Calabrese:2005in}. 
Nonetheless, systems hosting Quantum Many-Body Scars (QMBS) have been shown to evade the fast thermalization owing to the presence of these small subset of the finite energy density eigenstates (i.e. scar states) which break ergodicity. This leads to long lived quantum revivals and slow thermalization. Such phenomenon paves the way for interesting many-body physics beyond ground states or low-lying states and sparks a deeper debate about the hard criterion for the division between ETH-abiding and ETH-violating systems.
\medskip

Like the case for single particle scar systems having both ergodic and non-ergodic wavefunctions based on initial conditions, QMBS are also sensitive to the same. One can envision preparing QMBS by starting with explicitly ergodicity breaking initial conditions, leading to a much less rapid thermalization. This slow thermalization roots from the approximate partitioning of the Hilbert space between thermalizing and symmetry `protected' subspaces \cite{moudgalya2022hilbert}. These scar subsector of states could show periodic revivals of the initial wavefunction, coupled with a large autocorrelation value. It turns out that one needs to be rather specific about such initial conditions to be able to probe the whole arc of these special eigenstates. Explicitly, the strong dependence of relaxation dynamics on initial condition in the presence of QMBS states (as opposed to the MBL scenario \cite{HuseMBL, RMPAbanin}) is very significant.  Thus, it becomes interesting to explore what insights we may get from the dynamics of a system hosting QMBS, which is the central motivation of the present work.
\medskip

%quantify the time evolution of systems containing QMBS in their spectrum, and this is what we set out to do in the present work.
To this end, we focus primarily on the PXP model \cite{Sengupta2004PRB, Hosho2012PRA} which describes Rydberg atoms in the regime of strong Rydberg blockade. Pertinent in the context of experiment, PXP model has been paradigmatic in exploring many-body scar physics. This model is manifestly non-integrable and breaks all the Lie symmetries. However, starting from the N\'eel or $|\mathbb{Z}_2\rangle$ initial product state, a `tower' of states can be constructed which forms an approximate $su(2)$ multiplet. The existence of such approximate $su(2)$ invariant subspace within the Hilbert space leads to periodic revivals in the dynamics, when the system is globally quenched from the aforementioned initial state.  Motivated by such intriguing aspects, we work towards an understanding of \textit{quantum complexity} of spreading an initial state to the protected scar states in the Hilbert space. Often called as ``spread complexity," in its simplest avatar it is associated to the optimal spread of a particular state under time-evolution \cite{Balasubramanian:2022tpr}. %efficient quantum circuit that takes a reference state into the desired target state given a set of quantum gates. 
Recently, this has found applications in the context of quantum many-body theory \cite{Balasubramanian:2022tpr, https://doi.org/10.48550/arxiv.2208.08452, Caputa2022PRB, Caputa:2022yju, https://doi.org/10.48550/arxiv.2208.10520,Erdmenger:2023shk,Bhattacharya:2023zqt,Chattopadhyay:2023fob,Pal:2023yik}. %, scrambling \cite{Barbon:2019wsy,Sonner2019} and by extension, in Holographic theories \cite{} and the Black Hole information loss paradox \cite{}.  
The notion of quantum complexity comes from the theory of quantum computation. One way of quantifying the complexity associated with certain quantum states is based on Nielsen's approach  \cite{NL1,NL2,NL3}, which is associated with the construction of the most efficient quantum circuit that takes a reference state into the desired target state given a set of quantum gates. Motivated by certain conjectures coming from AdS/CFT correspondence, in particular, the conjecture that it resolves certain puzzles related to black holes \cite{susskind1,susskind2}, it has now been used for probing various aspects of quantum systems  \cite{Jefferson,Chapman:2017rqy,Bhattacharyya:2018wym,Caputa:2017yrh,me1,Bhattacharyya:2018bbv,Hackl:2018ptj,Khan:2018rzm,Camargo:2018eof,Ali:2018aon,Caputa:2018kdj,Guo:2018kzl,Bhattacharyya:2019kvj,Flory:2020eot,Erdmenger:2020sup,Ali:2019zcj,Bhattacharyya:2019txx,cosmology1,cosmology2, Caceres:2019pgf,Bhattacharyya:2020art,Liu_2020,Susskind:2020gnl,Chen:2020nlj,Czech:2017ryf,Chapman:2018hou,Geng:2019yxo,Guo:2020dsi,Couch:2021wsm,Erdmenger:2021wzc,Chagnet:2021uvi,Koch:2021tvp,Bhattacharyya:2022ren,Bhattacharyya:2023sjr,Bhattacharyya:2022rhm,Bhattacharyya:2021fii,Bhattacharyya:2020iic}\footnote{This list is by no means exhaustive. Interested readers are referred to these reviews \cite{Chapman:2021jbh,Bhattacharyya:2021cwf}, and references therein for more details.}.  
Our target in this work is to explore \textit{how a particular notion of quantum complexity, namely spread complexity, can be exploited as a diagnostic of scar eigenstates in many-body systems.}
\medskip

As it turns out, usually the computation of quantum complexity is highly basis dependent \cite{me1}, but one can show that the minimum is attained when the chosen basis is the so-called Krylov basis \cite{Balasubramanian:2022tpr}. The minimization of the spread complexity in the Krylov basis requires extracting the Lanczos coefficients associated to the Hamiltonian of the model \footnote{A related but different notion of complexity associated with operator growth in Krylov basis, often termed as \textit{K Complexity}, has been recently explored in the literature. Interested readers are referred to some of these references \cite{DymarskyPRB2020,PhysRevLett.124.206803,Yates2020,Yates:2021lrt,Yates:2021asz,Dymarsky:2021bjq,Noh2021,Trigueros:2021rwj,https://doi.org/10.48550/arxiv.2207.13603,Fan_2022,Kar_2022,Barbon:2019wsy,Rabinovici_2021,Rabinovici_2022,Rabinovici2022,https://doi.org/10.48550/arxiv.2109.03824,PhysRevE.106.014152,https://doi.org/10.48550/arxiv.2204.02250,Bhattacharjee_2022a,Du:2022ocp, Banerjee:2022ime, https://doi.org/10.48550/arxiv.2205.12815,https://doi.org/10.48550/arxiv.2207.05347,H_rnedal_2022,https://doi.org/10.48550/arxiv.2208.13362,Du:2022ocp,Bhattacharjee:2022lzy,Alishahiha:2022anw, Avdoshkin:2022xuw,Kundu:2023hbk,Rabinovici:2023yex,Zhang:2023wtr,Nizami:2023dkf} and references therein. This list of references by no means is exhaustive.}. 
A previous discussion in the literature \cite{Bhattacharjee_2022} attempted to approximate the broken $su(2)$ symmetry within the scar subspace of the parent Hamiltonian using a quantum-deformed $su(2)_q$ symmetry and associated Lanczos coefficients. However, in the present work we choose to take a slightly but significantly different path towards the same.
Instead of approximating, we embrace the breakage in symmetry for the PXP system via a rigorous implementation of the Forward Scattering Approximation (FSA), which points towards subtle differences in computing the Lanczos coefficients. Explicitly, Lanczos coefficients for a system size $L$ in the $|\mathbb{Z}_2\rangle$ state,  calculated from the FSA algorithm, swiftly closes after $L+1$ steps, however those from the vanilla Lanczos algorithm  continue to remain finite. Moreover, the FSA is more sensitive in accessing the scar subspace of the system by construction. We would take advantage of this subtlety to explicitly construct the spread complexity from different protected initial states, including  $|\mathbb{Z}_2\rangle$,  $|\mathbb{Z}_3\rangle$ and also the vacuum $|0\rangle$. The structure of scar subspaces built on top of these initial states vary significantly, with  $|\mathbb{Z}_2\rangle$ providing the strongest revivals, a fact we will observe in the case of spread complexity as well. Complexity starting from the $|\mathbb{Z}_3\rangle$ state would have weaker revivals, which we will show using an FSA-like paradigm. For both those cases, near-exact revivals can be restored using particular set of perturbations. For the vacuum $|0\rangle$ initial state, although there are no scar subspace present per se, judiciously adding appropriate few body terms to the bare Hamiltonian would still result in revivals. The imperfect nature of FSA is quantified by the so called ``FSA errors" which is found to have a rich structure and carry important insight about the scaling of the dynamics with systems size.
\medskip

The rest of the paper is organized in the following way: we start with an introduction to quantum scars in section \ref{sec1}, and move to an exposition of spread complexity of states in \eqref{sec2}. In section \eqref{sec3} we will have a brief look at the nature and structure of the PXP model, which we will be studying throughout this work. Section \eqref{sec4} contains the central discussions of the paper, wherein we describe the subtle differences between following the vanilla Lanczos method and the FSA for the PXP model. We also show how the spread complexity calculated from the $|\mathbb{Z}_2\rangle$ and $|\mathbb{Z}_3\rangle$ initial states evolve over time. In \eqref{sec5} we describe the more subtle case of $|0\rangle$ scars, which can be engineered by driving the PXP system with a square pulse protocol in a specific frequency range, and are not at all analogue of the $|\mathbb{Z}_2\rangle$ or $|\mathbb{Z}_3\rangle$ scars. Using Floquet perturbation theory one can write generic perturbations to the bare Hamiltonian that make such scar induced sustained oscillations possible, which we use to compute spread complexity for such initial states. We then wrap up our arguments and discuss future extensions in the section \eqref{sec6}. Appendices at the end contain additional computational details and discussions. 

\section{Scars and dynamical symmetry}
\label{sec1}
We have already discussed scars in continuum systems, in the sense that classically unstable periodic orbits of the stadium billiards scar, or, leave an imprint on a wavefunction. For quantum many-body systems, the telltale sign of scarring comes in the form of revivals of the wavefunction as it periodically returns to its value at an initial state. It has been argued \cite{Bull2020} that exact revivals in these models, or put another way, embedding of scar states in such a system comes from certain dynamical symmetries of the effective Hamiltonian \footnote{See \cite{barut1972dynamical} for an introduction to dynamical symmetry.}. These dynamical symmetry perspectives also point towards an underlying spectrum generating algebra, that gives rise to one or more towers of QMBS \cite{Khemani2020,Bull2020}. 
\medskip

The notion of scars as dynamical symmetries follow from the simple observation that there could exist a local operator $Q$ associated to (a part of) the system Hamiltonian such that 
\be{}
[H,Q] = \alpha Q, ~~\alpha \in \mathbb{R},
\ee
which immediately makes sure there are equidistant tower of states starting from a single eigenstate of $H$ that can be obtained by repeatedly acting with $Q$. In that case, one can think of $Q$ as the raising operator for states with this dynamical symmetry. Similarly one would have a lowering operator $Q^\dagger$ such that
\be{}
[H,Q^\dagger] = -\alpha Q, ~~\alpha \in \mathbb{R}.
\ee
If such a dynamical symmetry is embedded in the system, the quantum evolution equation $\dot{Q} = -i[H,Q]$ implies there would be persistent oscillation of constant frequency in this sector. These finite dimensional invariant subspaces will host the scar states, and the ladder operators would not be able to move them outside of the block. 
\medskip

Once we can exactly embed such scar states using the dynamical symmetry of the (partial) Hamiltonian, it has been shown \cite{Bull2020} that the total Hilbert space fragments into a thermalizing and a non-thermalizing block-diagonal part, with the scar states well-approximated by the basis vectors (or superposition of basis vectors) of a \textit{Krylov subspace}, where thermalization is absent \cite{moudgalya2022thermalization}. One can also visualise this as an integrable subspace detached from a fast thermalizing bulk. These Krylov subspaces are tri-diagonal bases optimal for time evolution of states, and we will come back to them in the next section in the context of complexity. One subtlety to note is that the embedding of the Krylov subspaces into the Hilbert space do not have to be exact for scar states to occur. The main system discussed in this work will be the PXP Hamiltonian, where it turns out the Krylov subspace only realizes a broken version of the $su(2)$ algebra, where one can try to restore the broken algebra to a very good extent by adding appropriate correction terms (both perturbative and non-perturbative) with very fine tuned sets of coupling strengths. Analytically, this depends on a clever algorithm to correct the root structure of the Lie algebra and mending broken commutation relations, resulting in strengthening of revivals \cite{Bull2020}.

\section{State complexity from Lanczos Algorithm and optimal spreading}\label{sec2}
In this section we will briefly review the steps behind the computation of state complexity from Lanczos Algorithm, which is also termed as \textit{Spread Complexity} \cite{Balasubramanian:2022tpr}. We start with the time evolution of a quantum state $|\psi(t=0)\rangle$ by a Hamiltonian $H.$ Then we have, 
\be{}
|\psi(t)\rangle= e^{-i\, H\, t}|\psi(t=0)\rangle=\sum_{k=0}^{\infty} \frac{(-i t)^k}{k !}\left|\psi_k\right\rangle,
\ee
where the last equality comes from the Taylor expanding the unitary $e^{-i\, H\, t}$ operator and defining, 
\be{}
H^{k}|\psi(t=0)\rangle=|\psi_k\rangle .
\ee
Although the set of $|\psi_k\rangle$ can be used to reconstruct the $|\psi(t)\rangle,$ the elements of $|\psi_k\rangle$ may not be orthonormal.  One can use a \textit{Gram-Schmidt} procedure to find another set of basis vectors $|K_n\rangle$ that does the job. These set of \textit{orthonormal} basis vectors $$\mathcal{K}=\Big\{|K_k\rangle \equiv v_k\,,\quad k=0,1,2,\cdots\Big\}$$ are commonly known as \textit{Krylov Basis}. This set of vectors ($v_k$) may not span the entire Hilbert-space but they span the the entire subspace accessible through the time-evolution of a given initial state $|\psi(t=0)\rangle.$ To generate these basis one typically uses the so called \textit{Lanczos Algorithm} \cite{SandvikREV}.
\medskip

To start the algorithm, one identifies $|K_0\rangle$ ($v_0$) with $|\psi (t=0)\rangle.$ Then the Lanczos algorithm provides us with the following recursion relation \cite{Parker_2019}, 
\be{}
\beta_{k+1}v_{k+1}= H v_k-\alpha_k v_k-\beta_k v_{k-1}, 
\ee
where, $\alpha_k=\langle K_k|H|K_k\rangle$ and $\beta_k=\langle A_k|A_k\rangle^{1/2}.$ 
%Also $|K_k\rangle$ is the normalized $|A_k\rangle$ i.e $|K_{k}\rangle=\frac{1}{\beta_k}|A_k\rangle.$ 
Also note that generically $\beta_0=0.$ %One can show that in the Krylov subspace the Hamiltonian takes the following form, 
%\be{}
%H=\left(\begin{array}{cccccc}
%\alpha_0 & \beta_1 & & & & \\
%\beta_1 & \alpha_1 & \beta_2 & & & \\
%& \beta_2 & \alpha_2 & \beta_3 & & \\
%& & \ddots & \ddots & \ddots & \\
%& & & \beta_{N-2} & \alpha_{N-2} & \beta_{N-1} \\
%& & & & \beta_{N-1} & \alpha_{N-1}
%\end{array}\right)
%\ee
%In fact for a $N \times N $ Hermitian matrix it is always possible to transform it in this a tri-diagonal or \textit{Hessenberg} form by doing a suitable similarity transformation of the form [],
%\be{}
%O. H. O^{T}= H_{\textit{Hessenberg}},\quad  O |\psi(t=0)\rangle= |\psi(t=0)\rangle. 
%\ee
%Here $$O=\left(\begin{array}{ccc}
%1 & 0 \\
%0 & M 
%\end{array}\right)$$ and $M$ is $N-1\times N-1$ matrix and the initial state which is always chosen to be the part of the Krylov Basis is the eigenvectors of $O$ with eigenvalue $1.$ \par
Given this Krylov basis we can define the \textit{Spread Complexity} in the following way \cite{Balasubramanian:2022tpr}, 
\be{eq1}
\mathcal{C}(t)=\sum_k k|\psi_k(t)|^2, 
\ee
where $\psi_k(t)=\langle \psi(t)|K_k\rangle$ is the basis coefficients when the time evolved state is expanded in terms of Krylov basis. One can verify that $\sum_k |\psi_k(t)|^2=1.$ This is consistent with the fact that the time evolution is unitary. In \cite{Balasubramanian:2022tpr}, it was shown that $\mathcal{C}(t)$ is optimal when the one computes it in the Krylov basis rather than using any other arbitrary basis. Hence the idea of spread complexity in Krylov basis can be thought of as some sort of measure quantifying the complexity of the spread of the wavefunction into the Hilbert space under time evolution. In the subsequent section we will use \eqref{eq1} to compute the spread complexity for PXP model with various choices of initial states. 
%\newpage

\section{PXP system: A lightning review}\label{sec3}

To begin with, let us provide the readers with a glimpse of the model we will be concerned with throughout. The Hamiltonian for the PXP model\footnote{Note that one can map the hamiltonian for an ultracold Rydberg atom chain to a PXP system in the blockade regime \cite{scar1}.} is given by:
\be{}
\mathcal{H}_{PXP} = \sum_{j=1}^{L}P_{j-1}\sigma^x_j P_{j+1},
\ee
    where $P_{j}=(|0\rangle\langle0|)_{j}$ is the projection operator at the $j-$th site to the ground state of $\sigma^{z}$ with $|0\rangle$ standing for the ground state of $\sigma^{z}$, and  $\sigma^{x} = |1\rangle\langle 0 + |0\rangle\langle 1|$ flips the state where $|1\rangle$ stands for the excited state of $\sigma^{z}$. This Hamiltonian possesses both translation and inversion invariance \cite{MotrunichPRL2019}. Translation invariance suggests one can diagonalize the above in a block diagonal form in a suitable momentum eigenbasis. Here, periodic boundary conditions are implied by identifying the $(L+1)$-th site to the $1$st site. The effective constraint on the system, as realised by the projection operators, suggests no two excited states can occupy two adjacent sites \footnote{States with neighboring excitations present would be eigenstates of the Hamiltonian with zero energy and appear to be frozen under dynamical evolution.}. The discrete spatial inversion symmetry of the system identifies sites $j  \to L-j+1$ \footnote{However if one considers Open Boundary Conditions, only inversion symmetry around the middle site is left.}. There is also a particle-hole symmetry, that maps $\mathcal{H}\to -\mathcal{H}$, i.e. maps eigenstates at $+E$ to a partner at $-E$.
\medskip

The kinematical constraint on the PXP system makes it non-integrable and fast thermalizing in the sense of level-spacing statistics. However, as we discussed, there are certain protected scar states in the spectrum that show violation of ETH. One can actually show that this model may be a deformation of an integrable Hamiltonian \cite{Khemani2020, Khemani2019, Sengupta2004PRB}, and hence the scar subspace could be thought of as a `memory' of the integrability. But in any case, majority of the states in the spectrum thermalize very fast, but states with a $|\mathbb{Z}_k\rangle$ symmetry thermalize slower than other non-symmetric states. One can show \cite{Turner2018} the $|\mathbb{Z}_2\rangle$ and $|\mathbb{Z}_3\rangle$ states are the most prominent special states which show weak thermalization behaviour. This nature has an inherent initial condition dependence,  and starting from a generic $|\mathbb{Z}_k\rangle$ state and following the evolution, one can show coherent oscillations occur in the quantum fidelity (or return probability) defined as:
\be{}
\mathcal{R}(t)=|\langle \psi(t=0)|\text{exp}(-i\mathcal{H}t)|\psi(t=0)\rangle|^2
\label{RP}
\ee
where $|\psi(t=0)\rangle$ denotes the initial state. Indeed, fidelity revivals for quenches from these states have been shown in the literature \cite{Turner2018}.
\medskip

Another very intriguing aspect of the PXP system is the broken $su(2)$ symmetry. A simple way to show that for the $|\mathbb{Z}_2\rangle$ initial state is to break the canonical Hamiltonian down into forward and backward propagating parts:
\begin{eqnarray}\label{hpm}
\mathcal{H}_{\pm} = \sum_{j\in \text{even}}P_{j-1}\sigma^{\pm}_j P_{j+1}+\sum_{j\in \text{odd}}P_{j-1}\sigma^{\mp}_j P_{j+1}
\end{eqnarray}
where $\sigma^{\pm} =\frac{1}{2}(\sigma^{x}\pm i\sigma^{y})$ and $\sigma$'s denote the usual Pauli matrices. These are generically called forward and backward scattering operators, because in an exact symmetric system they act as raising and lowering operators, and we can use them to write any element in the Hamiltonian matrix. However, in the case of PXP model, it turns out, they only approximately satisfy a $su(2)$ algebra. This makes life harder in the sense usual analytical techniques cannot be straightforwardly used to tabulate the spectrum. More specifically, for a perfectly $su(2)$ invariant system, starting from a lowest weight state, i.e. a simultaneous eigenstate of $\mathcal{H}_z = \frac{1}{2}[\mathcal{H}_+,\mathcal{H}_-]$ and the casimir operator, one would be able to create a tower of $su(2)$ invariant states. But in the case of PXP, due to the broken $su(2)$ algebra, a simple calculation yields:
\be{}
[\mathcal{H}_z, \mathcal{H}_\pm] = \pm \mathcal{H}_{\pm} +\Delta^{(1)}_{\pm},
\label{comm}
\ee
where the extra terms in the broken algebra can be written as:
\begin{align}
\begin{split}
    2\Delta_{\pm}^{(1)} = &\mp\sum_{j \in \text{even}} 
    \left(P_{j-2}P_{j - 1}\sigma^{\mp}_{j} P_{j + 1} +  P_{j-1}\sigma^{\mp}_{j} P_{j + 1}P_{j + 2}\right) \\&
    \mp \sum_{j \in \text{odd}}\left( P_{j-2}P_{j - 1}\sigma^{\pm}_{j} P_{j + 1} +  P_{j-1}\sigma^{\pm}_{j} P_{j + 1}P_{j + 2}\right)\,
\label{pertzz}
\end{split}
\end{align}
that is a string of projectors acting on consecutive neighbouring sites. It should be clear that adding these terms back to the Hamiltonian as a perturbation \cite{Bull2020} should improve the $su(2)$ representation by more accurately closing the commutators \footnote{Note that one could also write second order error-correcting terms added to the Hamiltonian. These are not higher order in $\lambda$, but include longer string of projection operators acting on neighbouring sites. In that logic $\lambda$ should not be considered a perturbation parameter in the usual sense.}. Combining Eqn. \ref{comm} and \ref{pertzz}, it is rather straightforward to show that the optimally perturbed Hamiltonian for $|\mathbb{Z}_{2}\rangle$ initial state can be written as
\begin{align}
\begin{split}
&\mathcal{H} = \mathcal{H}_{PXP} + \lambda \mathcal{H}_{P, \mathbb{Z}_{2}}, \\& 
\mathcal{H}_{P, \mathbb{Z}_{2}} = \sum_{j} (P_{j-2}P_{j-1}\sigma^{x}_{j}P_{j+1} + P_{j-2}\sigma^{x}_{j-1}P_{j}P_{j+1})
\label{PXPP}
\end{split}
\end{align}
We will come back to this point when we discuss complexity from the  $|\mathbb{Z}_2\rangle$ initial state. But we must caution the reader, the addition of this perturbation does not restore integrability to the PXP model 
\cite{Choi2019PRL}, merely improves the symmetry in a particular sector. For example, restoring the $su(2)$ in $|\mathbb{Z}_3\rangle$ sector requires a different set compatible raising and lowering operators, and hence a different set of perturbations. 

\section{Evaluating complexity for the PXP system}\label{sec4}

We now come down to the crux of this paper, where we discuss explicit computations of quantum complexity for both the unperturbed and optimally perturbed PXP model from different types of initial states. As is very well understood, since the fragmentation of the Hilbert space into thermal and protected regions is only approximate in the present case, we cannot directly use analytical procedures such as in \cite{Caputa2022PRB}. The corresponding Lanczos coefficients need to be calculated numerically. In this section, we will carefully define the method of Forward Scattering Approximation (FSA) and highlight the subtle difference between that and a blind Lanczos process. Later on, we will use these coefficients to compute the spread complexity from different initial states.
\medskip

\subsection{FSA vs Lanczos: Subtleties}

As we argued beforehand, the construction of special symmetry protected states in the PXP model comes about from a modification of the Lanczos algorithm. One considers Lanczos algorithm to project an approximation of the total Hamiltonian to the associated Krylov subspace. In the systems with scars, Krylov subspace is identified with the manifold where scar states live, albeit an exact Krylov diagonalization is not possible for the broken symmetry in the PXP case. To address this issue, one needs to resort to a non-exact version of FSA.
\medskip

Let's remind the reader that orthonormalizing the Krylov subspace requires to perform the following iteration:
\be{}
\beta_{n+1}v_{n+1}= H v_n-\alpha_n v_n -\beta_n v_{n-1},
\ee
i.e. the recursive application of the Hermitian Hamiltonian on the vector $v_n$, generates a set of orthonormal vectors. Note that the action of the Hamiltonian results in getting both the next and the former vector in the set, providing both forward and backward propagation of the vectors. The set of vectors $\{v_n\}$ is known as the Lanczos basis vectors, and the simple structure of the iterative algorithm guarantees global orthonormality. 
\medskip

\begin{figure}[ht!]
\includegraphics[width=8cm]{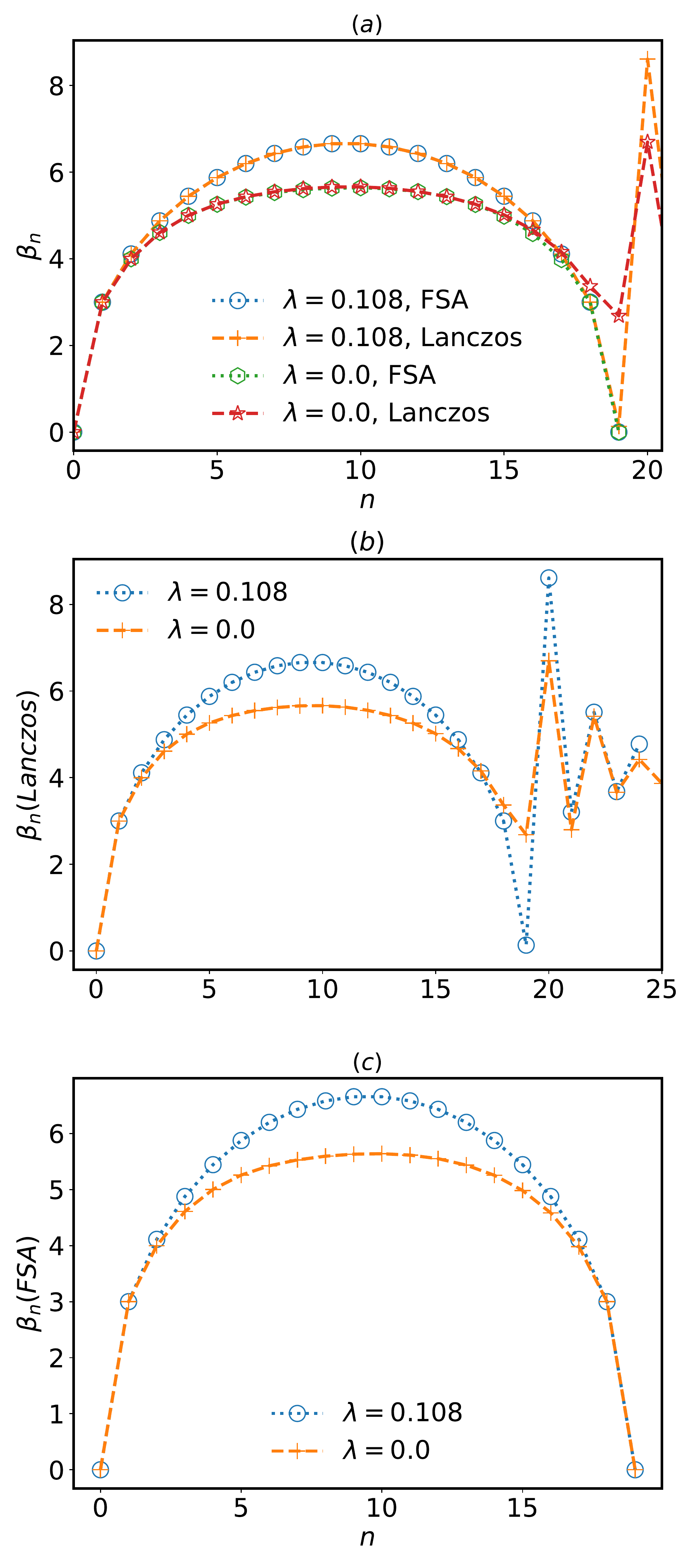}
\caption{Panel (b) and (c) shows $\{\beta_{n}\}$ coefficients (Note that the set $\{\alpha_{n}\}$ is identically 0) as a function of iteration number $n$,  respectively for the Lanczos and FSA method. The analysis is performed for $\mathcal{H} = \mathcal{H}_{PXP} + \lambda \mathcal{H}_{P, Z_2}$, with a system size of $L=18$. As clearly seen, the $\{\beta_{n}\}$ coefficients are identically 0 after $L+1$ steps whereas those for the straightforward Lanczos method are finite for any $n$ (even at the 19th step the value of $\b\sim 0.13$ for optimal perturbation); which is an important difference between FSA and Lanczos. However, irrespective of the value of $\lambda$, as long as one is interested only in the non-zero $\{\beta_{n}\}$ set for FSA, identical results are produced by Lanczos, as shown in panel (a).}
\label{fig:BCH_compare_b}
\end{figure}

The two-sided propagation for the Lanczos algorithm leads to the related scheme of FSA, where one can split the Hamiltonian in forward and backward scattering parts, like what we did in \eqref{hpm}. Once such a decomposition can be provided for a system with a global symmetry, we can in principle solve for the whole spectrum of the theory. In general this is completely compliant with the usual Lanczos technique. However when such a symmetry is only approximate, the naive Lanczos techniques do not close after $L+1$ steps, with $L$ being the system size, as the approximate Krylov subspace is not entirely disjoint from the thermalizing bulk. In this case, FSA for protected initial states in PXP model needs to be approached with caution, as the sheer requirement of convergence may not keep it equivalent to the case of simply following the naive Lanczos algorithm.\medskip

Let us now briefly illustrate the FSA method for the PXP model starting from a $|\mathbb{Z}_2\rangle$ initial state. As before, we consider the decomposition:
\be{}
\cH = \cH_{+} + \cH_-,
\ee
since the $|\mathbb{Z}_2\rangle$ state looks like $|101010...\rangle$, and hence $\cH_-$ directly annihilates it, while action of $\cH_+$ takes it to the next weighted vector $\b_1 v_1$. Since the vector $v_1$ is automatically orthogonal to $v_0$, we get $\a_0 = \langle Hv_0|v_0\rangle = 0$. Similarly for the second order recurrence, we would have,
\be{}
\beta_2 v_2 = \cH v_1-\b_1 v_0,~~\a_1 = 0.
\label{FSA_rec}
\ee
Computation using the decomposed Hamiltonian results in $\cH_- v_1 = \b_1 v_0$, $\cH_- v_2 = \b_2 v_1$
 etc, which combined with the content of Eqn. \ref{FSA_rec}, leads to a generic but compact recurrence formula:
\bea{}
\cH_+ v_n &=& \b_{n+1}v_{n+1} ~~\text{Forward Scattering}\label{for2}\\
\cH_- v_n &=& \b_k v_{n-1}~~\text{Backward Scattering}\label{back2}
\eea
and of course $\a_n = 0, \forall {n}$ in this case. One can clearly see that after $L+1$ iterations we get back to the $|\mathbb{Z}'_2\rangle =|0101010...\rangle$ state, which is promptly annihilated by $\cH_+$ and the process closes, unlike the naive Lanczos case. However, there is a catch; the above propagation equations are only valid for exact symmetries, for example in the case of the free paramagnet with $\cH = \sum_j \sigma^{x}_j$, where an exact $su(2)$ symmetry can be found\footnote{For exact $su(2)$ case the Lanczos coefficients are given by $\b_n = \sqrt{n(L-n+1)}$.}. For our case of interest, i.e. PXP system with $|\mathbb{Z}_2\rangle$ ($\ket{0}$ as well) inital state, the backward propagation becomes non-exact after two iterations of the algorithm due to the approximate symmetry. One then needs to define the FSA error by 
\be{fsaerr}
\delta v_n = ||\cH_- v_n - \b_n v_{n-1}||,~~n>2.
\ee
It has been shown in \cite{scar1} that the error in individual steps of FSA iterations can be thought of as a leakage of many-body wave functions outside of FSA manifold and can be related to the commutator of forward and backward scattering Hamiltonians. We do a detailed comparative analysis of these FSA errors in Appendix.~\ref{apA}, more specifically, we analytically calculate the FSA error in the first non-trivial FSA step which is found to display a non-monotonic (monotonic) behavior with system size for the $\mathbb{Z}_2$ ($\ket{0}$) state. We numerically find the higher FSA steps to also follow the similar trends. We argue that the scaling of the FSA errors can give important insight on the scaling of the dynamics with increasing system size.
\medskip

Following through with forward scattering algorithm, we can find the effective Hamiltonian projected onto Krylov subspace takes the form of a $(L+1)\times(L+1)$ tri-diagonal matrix:
\begin{equation}
  \label{eq:tmatrix}
  H_\text{FSA} = \left(\begin{array}{ccccc}
    0 & \beta_1  & & \\
    \beta_1  & 0 & \beta_2  & \\
             & \beta_2  & 0 & \ddots \\
             &          & \ddots   & \ddots  & \beta_{L} \\
             &          &          & \beta_{L} & 0
   \end{array}\right)\text{.}
\end{equation} 
Now, these Lanczos coefficients for PXP do not have closed analytic expressions \footnote{For their comparison with $su(2)_q$ Lanczos coeffiecients, see \cite{Bhattacharjee_2022}}, but in Fig \eqref{fig:BCH_compare_b} we show a detailed comparison for ${\b_n}$ numerically computed using both the naive Lanczos algorithm and the accurate FSA method for the context under consideration. It can be clearly observed that while simple Lanczos algorithm gives rise to coefficients which remain finite for any number of iterations, FSA set of ${\b_n}$ goes exactly to zero after $L+1$ iterations. We also show the effect of adding the perturbation we discussed in \eqref{pertzz}, since this should approximately close the Lie algebra and confine the dynamics in the scar manifold which results in undamped oscillations. In both cases adding the optimal perturbation (with $\lambda = 0.108$) almost gives rise to explicit $su(2)$ Lanczos values, but absolute convergence still happens only for the FSA case. Such conclusion holds for any system size $L$.

\subsection{Complexity and revivals from $|\mathbb{Z}_2\rangle$ state}
\begin{figure*}
\includegraphics[width=\textwidth]{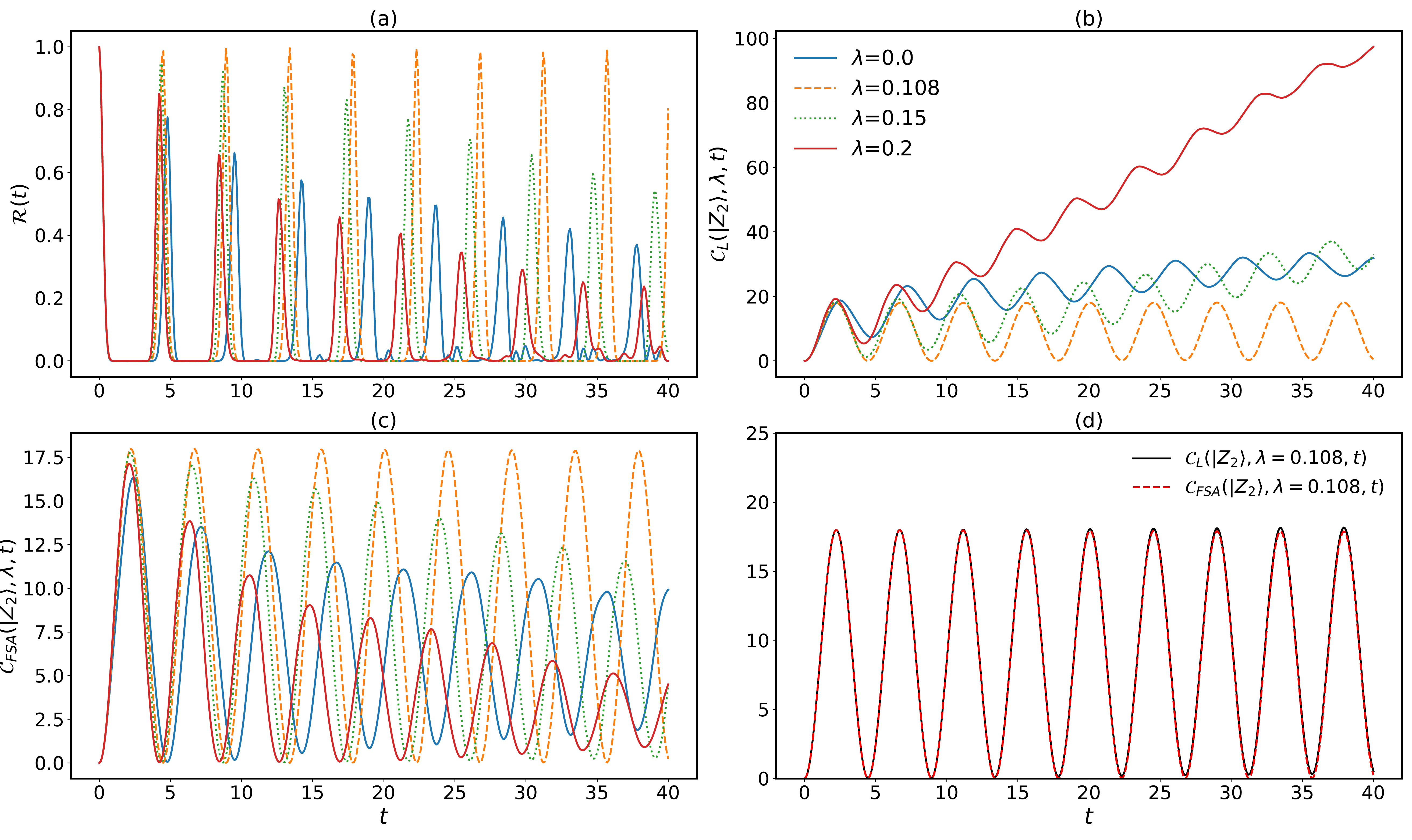}
\caption{Panel (a) shows the return probability $\mathcal{R}(t)$ defined in Eqn. \ref{RP} as a function of $t$ for different values of perturbation strengths $\lambda$ for the system $\mathcal{H} = \mathcal{H}_{PXP} + \lambda \mathcal{H}_{P, \mathbb{Z}_2}$, with $|\mathbb{Z}_2\rangle$ initial state. We see that revival is strengthened at/near a very special value of $\lambda=0.108$, which signifies the best possible restoration of the $su(2)$ symmetry at the aforementioned point. Larger $\lambda$ ($\gtrsim 0.15$) can significantly destroy the revivals. Panel (b) shows the quantum complexity $\mathcal{C}_{L}(|\mathbb{Z}_2\rangle, \lambda, t)$ computed via Lanczos method. At $\lambda=0$ i.e., for the unperturbed PXP model the (imperfect) revivals in $\mathcal{R}(t)$ are also captured via $\mathcal C_{L}(t)$. 
%The imperfect revivals are reflected in $\mathcal{C}_{L}(t)$ via weak change in oscillation amplitude and its envelop. 
The destruction of revival for larger $\lambda$'s are also reflected via $\mathcal{C}_{L}(t)$. In panel (c), we show the complexity $\mathcal{C}_{FSA}(|\mathbb{Z}_2\rangle, \lambda, t)$ computed from FSA. Not only does it capture perfect and imperfect revivals and the destruction of the same, but also the the nature of oscillation has rather striking qualitative similarity with that for $\mathcal{R}(t)$. Finally in panel (d), we demonstrate $\mathcal{C}_{L}(t)$ and $\mathcal{C}_{FSA}(t)$ for the special perturbation strength $\lambda=0.108$, which restores the $su(2)$ symmetry to a very good extent within scar subspace leading to perfect revivals.} 
\label{fig_BCH_compare}
\end{figure*}

Now that we have our methods in place, let us move on to discuss the complexity computed from various symmetry protected initial states using \eqref{eq1}, most important of them being the $|\mathbb{Z}_2\rangle$. In our set-up, as we described, we expect revivals in the return probability $\mathcal{R}(t)$ to happen periodically as the subspace of states go from $|\mathbb{Z}_2\rangle$ to $|\mathbb{Z}'_2\rangle$. Of course, since the symmetry of the Krylov subspace is only approximate, we would expect the strength of revivals to go down over time, and that is exactly what we see in Fig \ref{fig_BCH_compare}, panel (a) for the pure PXP time evolution. Adding perturbations of the form \eqref{PXPP} remedies the situation, especially when $\lambda = 0.108$ (system size $L=18$) , we can get pristine revivals in return probability. As expected, larger $\lambda$ destroys the pristine revivals again. In such regime of $\lambda$, as we will show later in Appendix \eqref{apB}, the system starts losing its otherwise present approximate $su(2)$ symmetry, and this departure cannot even be captured via  a $su(2)_q$ approximation. 
\medskip

In Fig \ref{fig_BCH_compare}, panel (b) we plot the spread complexity $\mathcal{C}_{L}(|\mathbb{Z}_2\rangle)$ associated to the scar states built on $|\mathbb{Z}_2\rangle$, which we will call as Lanczos complexity in the rest of the paper to avoid confusion. Now, we should remember the Lanczos coefficients calculated from the naive algorithm resulted in a failure of closure, i.e. the coefficients never went to zero, so naturally for complexity a deviation from perfect revivals is expected. In fact, the complexity in this case does not come back to the $|\mathbb{Z}_2\rangle$ value as it goes beyond the finite number of basis states we ought to have been working with. This situation is again corrected by the inclusion of the appropriate perturbation \eqref{PXPP}, which almost exactly stabilises the revivals of complexity for $\lambda = 0.108$, as shown in \cite{Choi2019PRL, Bull2020}. For higher values of the perturbation parameter, the periodic revivals are progressively washed away as the approximate symmetry gets destroyed, and the resulting Lanczos complexity just keeps increasing within the time window of our observation in which corresponding $\mathcal{R}(t)$ is already damped out. 
\medskip

Since FSA provides a better handle over the scar subspace dynamics, the complexity plotted in Fig. \ref{fig_BCH_compare}, panel (c) has substantially different structures. Specifically, the pure PXP revivals are much more robust than its naive Lanczos counterpart, albeit they also decay over time owing to the approximated method we use resulting in leakage. While adding the special perturbation restores almost pristine revivals, other values of $\lambda$ do not immediately destroy the revival characteristics. One can note, FSA complexity maximas almost coincide with fidelity minimas, since spread of the wavefunction is supposed to be highest at those points where autocorrelations values are lowest. Importantly, the complexity $\mathcal{C}_{\text{FSA}}$ remains bounded (unlike $\mathcal{C}_{L}$) for any of these situations within the time scale of our observation. Nonetheless, it is quite intriguing that at the special symmetry restoring point $\lambda=0.108$ for the perturbation described in Eqn. \ref{PXPP}, FSA complexity as compared to naive Lanczos computation shows almost same numbers. 
%\sn{\{}But we have to remember, this was the point where $\{\b_n\}$ calculated from FSA and those from approximately closed Lanczos algorithm would have almost agreed, as seen in Fig. \eqref{fig:BCH_compare_b}. We will see in later sections that it is not a generic trend since finding exact symmetry restoration points become considerably harder for other initial states.\sn{\}. [I checked and they are not !]} 

\subsection{Towards complexity for the $|\mathbb{Z}_3\rangle$ state}

As we discussed earlier, PXP model is known to show revivals starting from other symmetry protected initial states as well.
In this section, we consider quench from the $|\mathbb{Z}_{3}\rangle = |100100...\rangle$ state and study subsequent behaviour of state complexity.
In contrast to $|\mathbb{Z}_{2}\rangle$, this one is a three-site configuration where the other states are found by translating the up spin position. Numerically it is known that revivals for these states decay very fast, indicating the ``brokenness'' of the underlying $su(2)$ symmetry algebra is even stronger (compared to the $|\mathbb{Z}_{2}\rangle$ case) for such states.
\medskip

\begin{figure*}
\includegraphics[width=\textwidth]{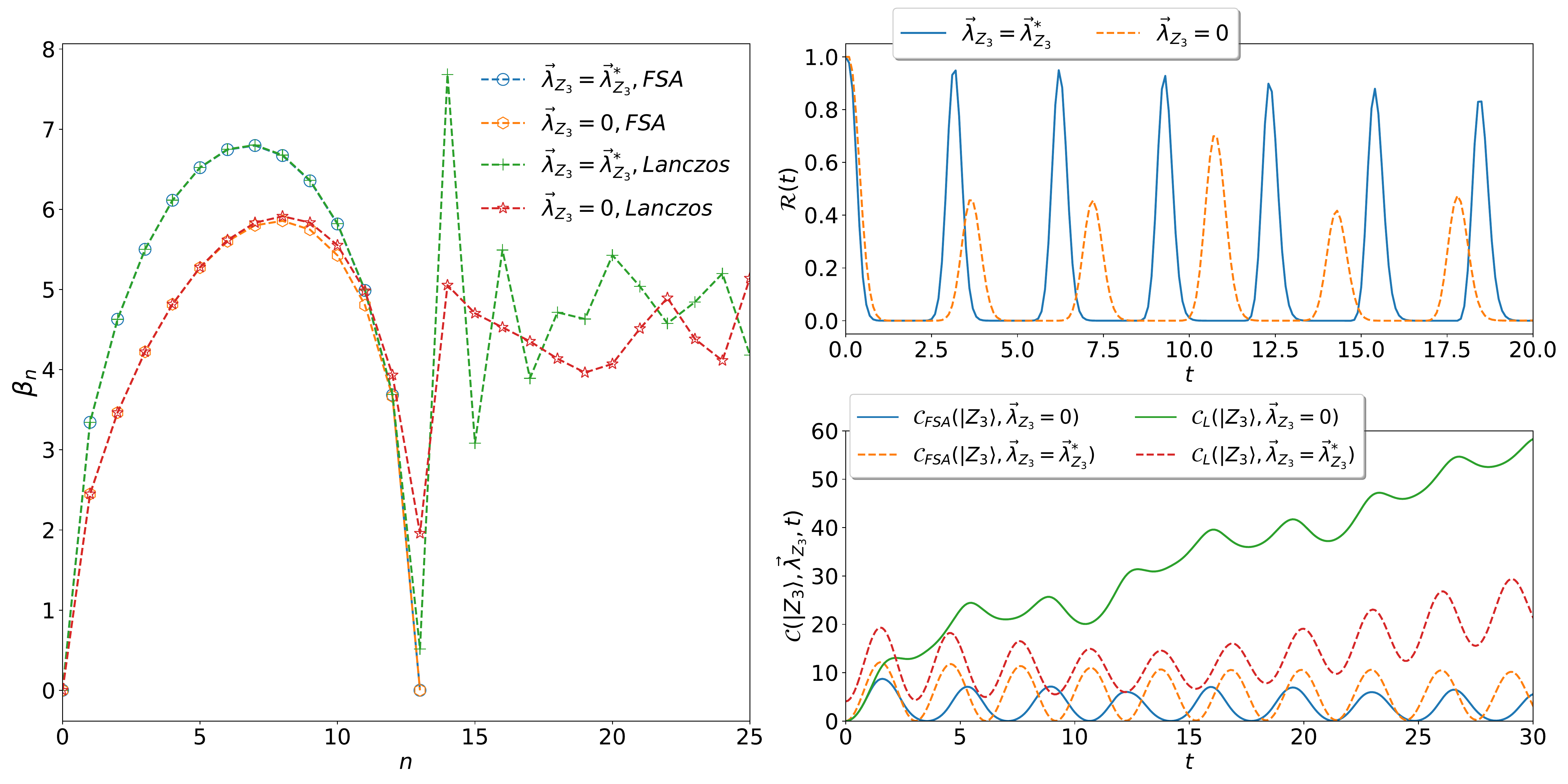}
\caption{Comparison of FSA-like and Lanczos methods for $|\mathbb{Z}_3\rangle$ case are shown in the left panel. $\vec{\lambda} = \vec{\lambda}^{*}_{\mathbb{Z}_{3}}$ denotes optimal symmetry restoring values of the perturbations given in \eqref{z3pertstrength}. Coefficients calculated from FSA and vanilla Lanczos at this value (blue and green dotted lines) almost coincide, however only the FSA curve closes at $2L/3$. The associated revivals for fidelity and complexity are plotted on the right hand panel. Note that the bare model shows a beating pattern on top of the revivals in $\mathcal{R}(t)$, indicating a modulation in oscillation amplitudes, this is however absent in the perturbed model.   %We should stress that it more accurate to call our procedure as `FSA-like'. 
Certainly, FSA complexity shares more structural similarity with $\mathcal{R}(t)$ compared to Lanczos complexity.}
\label{fig:z3summary}
\end{figure*}

Even though FSA is not properly defined for such an initial state, and consequently proper closed analytic forms of perturbations to stabilise oscillations are not clearly known, we proceed with the following general approach (as advocated in \cite{Bull2020}) and start by introducing the raising and lowering operators compatible with $|\mathbb{Z}_{3}\rangle$ state:
\begin{align}
    \bar{\cH}_{+} = \sum_{j}\bigg(\tilde{\sigma}^{-}_{3j} + \tilde{\sigma}^{+}_{3j+1} + \tilde{\sigma}^{+}_{3j+2}\bigg) \\
    \bar{\cH}_{-} = \sum_{j}\bigg(\tilde{\sigma}^{+}_{3j} + \tilde{\sigma}^{-}_{3j+1} + \tilde{\sigma}^{-}_{3j+2}\bigg)
\end{align}
where we have adopted the shorthand $\tilde\s_j = P_{j-1}\s_j P_{j+1}$ for brevity, and will stick to this notation in the rest of the paper. Given such decomposition of PXP Hamiltonian, one can proceed from here in a FSA-like method \footnote{This cannot really be termed as a proper FSA since the end state after a $2L/3$ cycle would be a superposition of states. However, for our purpose, it gives excellent results. } which terminates after $\frac{2L}{3}+1$ steps where $L$ is the system size. Note that this is a distinct scar subspace compared to the one built atop $|\mathbb{Z}_2\rangle$ initial state, and therefore contains a different representation of broken $su(2)$ symmetry. Similar to the $|\mathbb{Z}_2\rangle$  case, the naive Lanczos method does not truncate in this case as well, we show this in the left panel of Fig. \eqref{fig:z3summary}. The FSA-like recursion method we use however closes after the exact number of steps and gives the proper closed set of Lanczos coefficients in this case. 
\medskip 

\paragraph*{Perturbations leading to revival}:
Following what we did for our discussion regarding $|\mathbb{Z}_2\rangle$ initial states in section \eqref{sec3}, we can check explicitly how broken is the symmetry algebra associated to the present case.
Defining the diagonal element as $\mathcal{\bar{H}}_z = \frac{1}{2}[\mathcal{\bar{H}}^+,\mathcal{\bar{H}}^-]$, one can see that the broken algebra takes the form we encountered before: $[\mathcal{\bar{H}}_z, \mathcal{\bar{H}}_\pm] = \pm \mathcal{\bar{H}}_{\pm} +\bar\Delta^{(1)}_{\pm}$, where the last term on the RHS is a collection of corrective terms. As calculated already in \cite{Bull2020}, we chart out the perturbations that revive oscillations from $|\mathbb{Z}_{3}\rangle$ state.
The set of perturbations to the Hamiltonian can be written as: 
\be{}
    \mathcal{H} = \mathcal{H}_{PXP}
    + \vec{\lambda}_{\mathbb{Z}_{3}}\cdot\ \vec{\mathcal{H}}_{P, \mathbb{Z}_{3}}'
\ee
where the perturbations contained as the components of  $\vec{\mathcal{H}}_{P, \mathbb{Z}_{3}}'$ are given by
\begin{widetext}
\begin{align}\label{pertz3}
    &\vec{\mathcal{H}}_{P, \mathbb{Z}_{3}}'^{(1)} = \sum_{n}\bigg(P_{3n-2}\sigma^{x}_{3n-1}P_{3n}P_{3n+1}+P_{3n-1}P_{3n}\sigma^{x}_{3n+1}P_{3n+2}+P_{3n-1}\sigma_{3n}^{x}P_{3n+1}P_{3n+2}+P_{3n-2}P_{3n-1}\sigma^{x}_{3n}P_{3n+1}\bigg)\nonumber,\\
    &\vec{\mathcal{H}}_{P, \mathbb{Z}_{3}}'^{(2)} = \sum_{n}\bigg(P_{3n}P_{3n+1}\sigma^{x}_{3n+2}P_{3n+3}+P_{3n}\sigma^{x}_{3n+1}P_{3n+2}P_{3n+3}\bigg)\nonumber, \\
    &\vec{\mathcal{H}}_{P, \mathbb{Z}_{3}}'^{(3)} = \sum_{n}\bigg(P_{3n}\sigma^{x}_{3n+1}\sigma^{x}_{3n+2}\sigma^{x}_{3n+3}P_{3n+4}+P_{3n-1}\sigma^{x}_{3n}\sigma^{x}_{3n+1}\sigma^{x}_{3n+2}P_{3n+3} \bigg)
\end{align}
\end{widetext}
Notice that the symmetry breaking terms here are manifestly different from what we encountered in the $|\mathbb{Z}_{2}\rangle$ case, with few-body terms having support over more sites. 
Revival of oscillations i.e., restoration of $su(2)$ symmetry to the best extent occurs at the special coupling strength 
\be{z3pertstrength}
\vec\lambda^*_{\mathbb{Z}_{3}}=(0.18244, -0.10390, 0.05445),
\ee
which was found using numerical optimization in \cite{Bull2020}. In fact, the fidelity from the $|\mathbb{Z}_3\rangle$ state using the perturbed hamiltonian shows almost pristine revivals (see Fig. \eqref{fig:z3summary})with these parameters. Clearly these are only first order perturbations, and the numerical optimizations over many terms is less exact, so eventually the revivals start to decay over time.
\medskip

\paragraph*{Complexity}:
Once we have the structures in place, we go ahead and compute the spread complexity exactly as in the $|\mathbb{Z}_2\rangle$ case. In Fig. \eqref{fig:z3summary} we compare between spread complexity computed both from FSA-like and vanilla Lanczos methods. Clearly the complexity calculated using the Lanczos method from bare PXP model again shows a growth within the time window of observation, even more prominently than that in the  $|\mathbb{Z}_2\rangle$ case. However, in the present case, adding the perturbations \eqref{pertz3} with optimum strengths do not completely stabilize the oscillations. On the other hand, complexity calculated from the FSA-like method already shows robust sustained oscillations in the bare case, only to be augmented by adding symmetry restoring perturbations. 

\subsection{A curious system with exact symmetry}

\begin{figure}[ht!]
\includegraphics[width=\columnwidth]{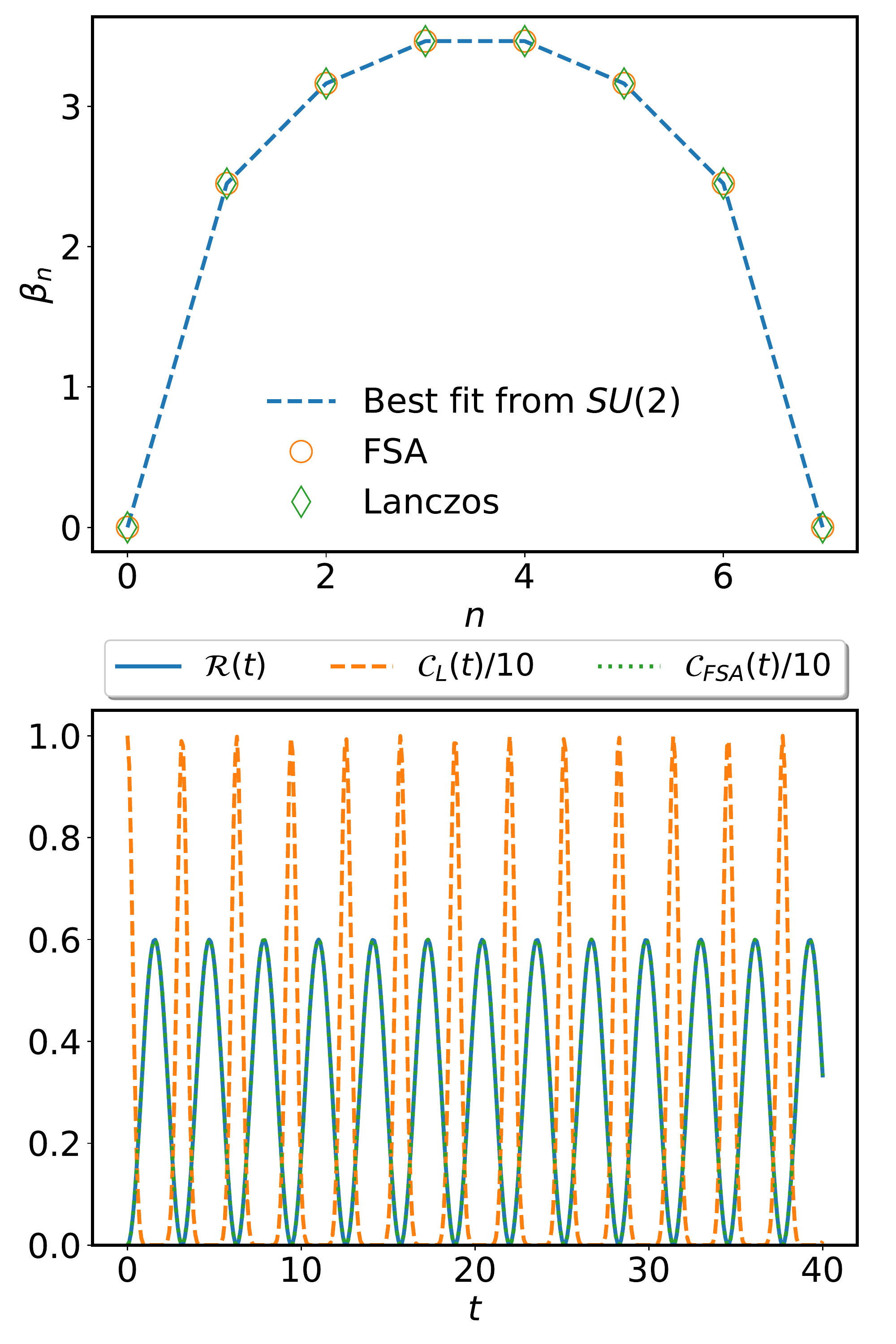}
\caption{Lanczos coefficients (upper panel) and complexity calculated for the system described in \eqref{exactz3}. Note that FSA closes after $L/3 +1$ steps (with system size $L=18$). The \textit{exact} $su(2)$ symmetry is very apparent in both these figures. } \label{H-V1}
\end{figure}

There exists a very curious feature of the $|\mathbb{Z}_3\rangle$ revivals, in the sense the symmetry in this subsector can be made exact \cite{Bull2020} by defining:
\be{exactz3}
\tilde{\cH} = \cH_{PXP}-V_1,
\ee
where the shift $V_1$ was defined previously in \eqref{pertz3}, and since the strength of the term here is of order unity, we prefer not to call it a perturbation. Sometimes this is termed as a strong deformation limit for the PXP model. For this case, one can still write down a FSA-like decomposition, where
\bea{}
\tilde{\cH}^{+} &=& \sum_{j}\bigg(\mathbb{I}-(P_{3j-2}+P_{3j+2})\tilde{\sigma}^{-}_{3j} \\ \nonumber &+&(\mathbb{I}-P_{3j-1}) \tilde{\sigma}^{+}_{3j+1} + (\mathbb{I}-P_{3j-4})\tilde{\sigma}^{+}_{3j+2}\bigg), 
\eea
and $\tilde\cH^{-}$ can be written similarly such that $\tilde\cH = \tilde\cH^{-}+ \tilde\cH^{+}$. This decomposition guarantees that the $su(2)$ algebra is exact in this sector for this Hamiltonian. However, since the action of $\bar{\cH}_{+}$ on the $|\mathbb{Z}_3\rangle$ state is just equivalent to spin flips at the $3j$ positions, the recursion repeats after $L/3$ steps, a marked difference from the earlier case. 
\medskip

We can numerically verify the Lanczos coefficients calculated from both FSA and naive Lanczos method in this case precisely follows $su(2)$ values, as we have shown in
Fig. \eqref{H-V1} upper panel. Similarly we plot the complexity calculated from these coefficients in the lower panel of the same figure, where naturally both $\mathcal{C}_L$ and $\mathcal{C}_{FSA}$ exactly coincide. In \cite{Bull2020} it has been shown that this exact restoration follows from a dynamical symmetry preserving unitary transformation on the algebra, brought about by the action of $V_1$.

\section{Complexity from the vacuum: A closer look}\label{sec5}

In general, one might think of the vacuum $|0\rangle$ as a generic product state that does not show any periodic revivals over time. But since we have been talking about adding perturbations that stabilize scar-induced revivals, one might ask whether such counterterms are present for the vacuum initial state as well. In \cite{BhaskarPRB2020}, such counterterms have been discussed in a Rydberg chain setting and the interesting features pertaining to the dynamics have been charted out. However, the system in question in this case is under a periodic drive with a tuning frequency, and resulting weak ETH violating oscillations could only coexist with the usual $|\mathbb{Z}_2\rangle$ scars in a particular drive frequency range \footnote{One could interpret these as Floquet Scars. For well defined Floquet protocols, one may be able to generate scarred dynamics from some initial states in the sense of a regular stroboscopic Floquet evolution.}. These scars are also non-uniform, much more fragile and thermalize faster than their $|\mathbb{Z}_2\rangle$ cousins (but still slower than bulk states). 
\begin{figure}[ht!]
\includegraphics[width=\columnwidth]{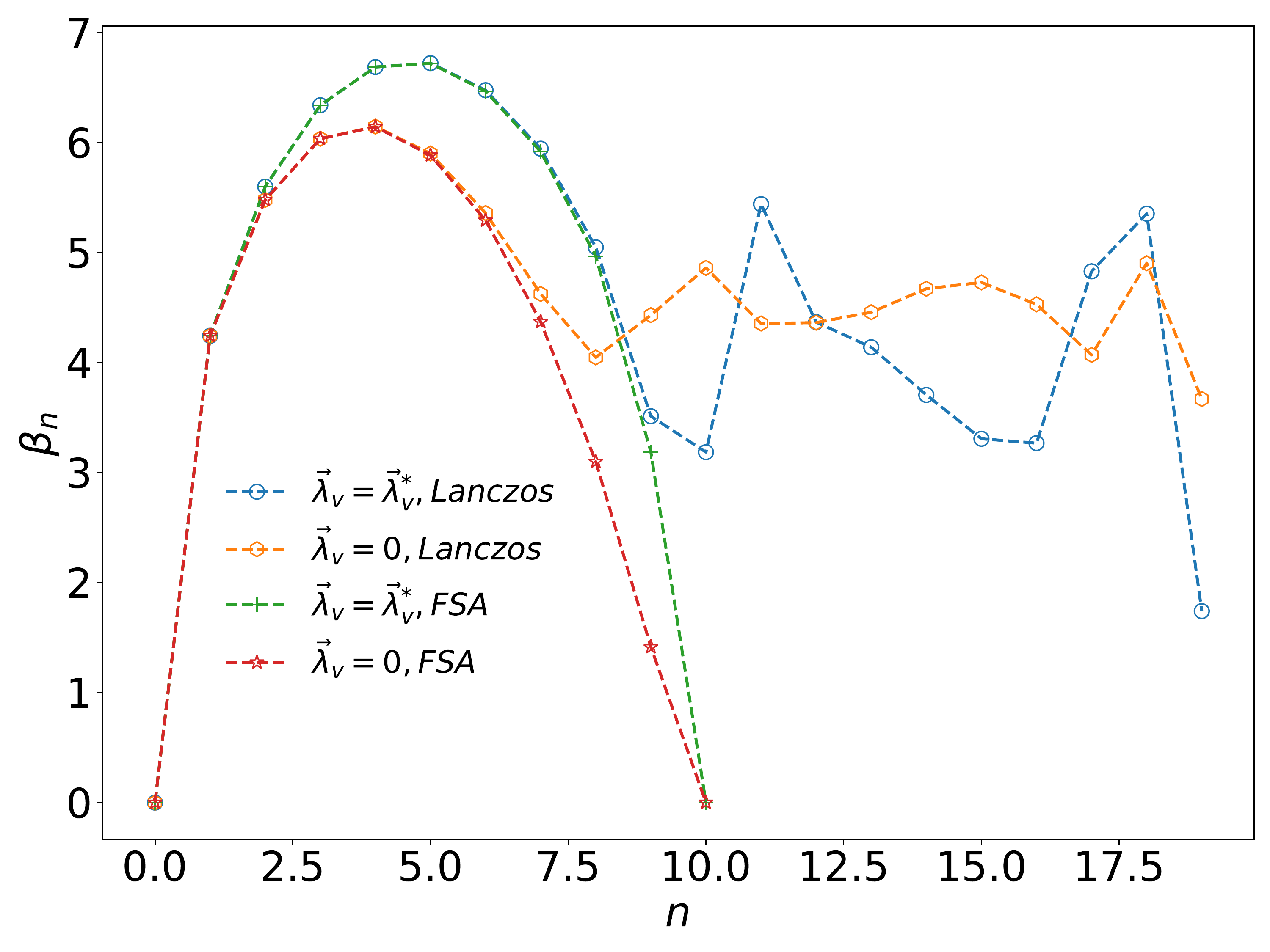}
\caption{Lanczos and FSA from the vacuum state. As seen before for $|\mathbb{Z}_2\rangle$ case, naive Lanczos from PXP does not close, while FSA closes after $\frac{L}{2}$ iterations. Adding a set of perturbative higher spin terms approximately restore the $su(2)$ symmetry to the system.} 
%\textcolor{red}{What is the system size here?} }
\label{fig:Similarity}
\end{figure}

\subsection{FSA from the vacuum state}
For vacuum initial state, the PXP Hamiltonian needs to be decomposed in a different way compared to the previous cases. We can write the decomposition as:
\be{}
 \cH_{PXP}=\cH_{+}+\cH_{-}=\sum_i(\tilde{\sigma}^{+}_i+\tilde{\sigma}^{-}_i).
\ee
Note the contrast between the above and \eqref{hpm} where the decomposition resulted in an even-odd pair. To motivate this choice, note that the action of $\cH_-$ on the vacuum $|00000...\rangle$ annihilates it and action of $\cH_+$ annihilates the final state, which in this case is $|\mathbb{Z}_2\rangle$ or $|\mathbb{Z}_2'\rangle$. The Krylov space in this case is spanned by $\frac{L}{2}+1$ vectors, as one can reach the final state(s) by acting on the vacuum  with $\cH_+$ a $L/2$ number of times. So likewise for the $|\mathbb{Z}_2\rangle$ initial state, this process closes exactly, very much in contrast to the naive Lanczos operation. In Fig.\eqref{fig:Similarity} we show a comparison between the two processes in calculating Lanczos coefficients. 
\medskip

As we discussed earlier, there is no analogue of scar states starting from the vacuum in the pure PXP case. It has been demonstrated in \cite{BhaskarPRB2020} that one needs few body terms in the Hamiltonian to stabilise oscillations from these scars. One could consider an effective Hamiltonian for this purpose: 

\begin{equation}\label{effvacH}
 \cH_{eff}=\sum_{i}\tilde{\sigma}^x_i+\lambda_3\sum_{i}\tilde{\sigma}^x_i\tilde{\sigma}^x_{i-1}\tilde{\sigma}^x_{i+1},
\end{equation}
where we have added three-site terms to the bare Hamiltonian. It could be easily seen that the effective decomposition in this case gives rise to:
\begin{equation}
 \cH^-_{eff}=\sum_{i}\tilde{\sigma}^-_i+\lambda_3\sum_{i}\tilde{\sigma}^+_i\tilde{\sigma}^-_{i-1}\tilde{\sigma}^-_{i+1}
\end{equation}
and $H^+_{eff}=(H^-_{eff})^T$ is the forward scattering counterpart of the same, with $\lambda_3$ being a real parameter. The effect of adding the 3-spin term is profound in this case as it can already create stable oscillations in the fidelity spectrum which was almost absent for the bare case. 

\begin{figure*}[ht!]
\includegraphics[width=\textwidth]{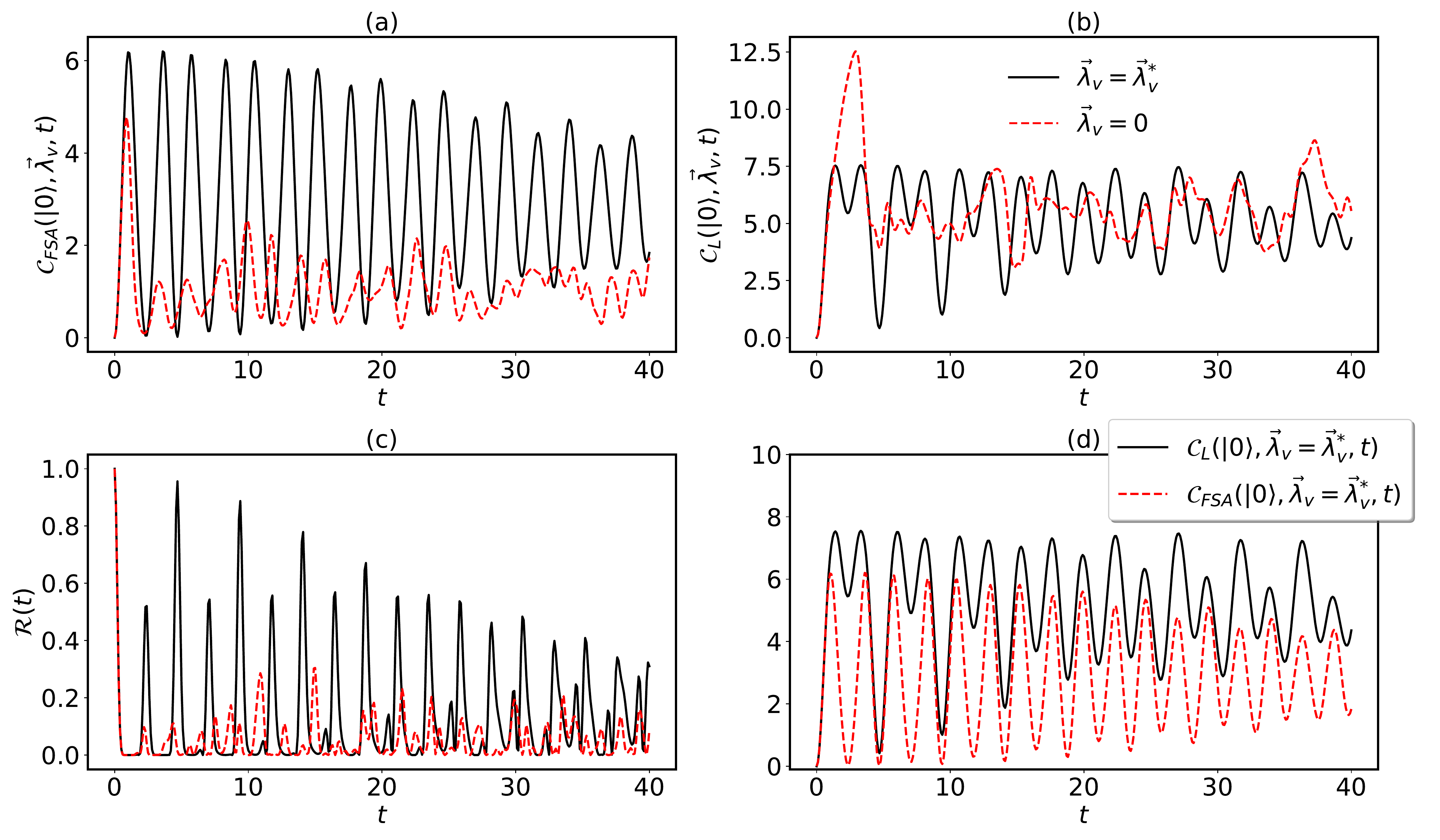}\caption{In panel (a) we show the complexity $\mathcal{C}_{FSA}(|0\rangle, \vec{\lambda}_{v}, t)$ calculated from the FSA for the vacuum state. Clearly, in absence of any perturbations i.e., $\vec{\lambda_{v}} = 0$ complexity damps down quite quickly as expected. However appropriate perturbations with corresponding appropriate strengths (see main text for details), we observe strong revivals in complexity. Panel (b) represents the same scenario where the complexity is computed via the naive Lanczos method. However, as it can be seen rather clearly, $\mathcal{C}_{L}(|0\rangle, \vec{\lambda}_{v} = \vec{\lambda}_{v}^{*}, t)$ has more substructures in oscillations which are absent in the return probability $\mathcal{R}(t)$, demonstrated in panel (c). In such a sense, the complexity computed via FSA shares somewhat more structural similarity with $\mathcal{R}(t)$, as opposed to that computed via Lanczos. Finally, panel (d) compares the complexity calculated from Lanczos and FSA at the point with optimal perturbation strength $\vec{\lambda}_{v}=\vec{\lambda}_{v}^{*}$ that to a good extent restores the otherwise broken $su(2)$ symmetry.} 
\label{fig:vacstate}
\end{figure*}

However, one must remember, as we discussed before, that adding this perturbation does not exactly restore the $su(2)$ symmetry. Hence the backward evolution step for FSA \eqref{back2} is still not exact for the system. In fact, one can show, in this case also, the non-trivial backscattering error starts from the third step of the FSA. To counter this, a minimization of the error is necessary, and the addition of the perturbation in \eqref{effvacH} gives one a handle in terms of the parameter $\lambda_3$. As will be briefly shown in Appendix \eqref{apA}, this error is dependent on the system size $L$ as well. In \cite{BhaskarPRB2020} it has been argued how errors at each non-trivial step of the FSA can be minimized and consequently robust oscillations can be generated by consecutively adding few body terms to the bare Hamiltonian. Semi-analytical arguments from Floquet perturbation theory dictates these extra terms are of the form $(2k+1)\tilde\s$ having support over $2k+3$ sites. For example, one can add a 5-spin term to the $\cH_{eff}$: 
\be{}
\cH = \cH_{eff}+ \lambda_5\sum_{j}\tilde{\sigma}^x_{j-1}\tilde{\sigma}^x_{j+1}\tilde{\sigma}^x_{j+2}\tilde{\sigma}^x_{j}\tilde{\sigma}^x_{j-2},
\ee
which one could again decompose into forward and backward moving ladder operators and get Lanczos coefficients. Then the minimization of FSA errors (i.e. maximization of oscillations) becomes the process of numerically finding optimal values of $\vec{\lambda}_{v}=(\lambda_3,\lambda_5)$.  This parameter space keeps increasing as we consecutively add more correction terms. 

\subsection{Results for Complexity from vacuum}

To compute spread complexity starting with the vacuum, we could start from the bare PXP model. However, as shown in Fig.\eqref{fig:vacstate}, autocorrelations from the bare model do not show any discernible revivals, which is also true for the complexity calculated from both naive Lanczos and FSA method. To get optimal revivals, we consecutively add $(2k+1)\tilde\s$ terms to the bare Hamiltonian upto $k=6$. We then fine tune the values of the parameters $\vec{\lambda}_{v}$ to get maximal oscillations in return probabilities from the vacuum. A very fine-tuned, albeit optimal set of solutions turns out to be the choice $\vec{\lambda^*_v}$: \{$\lambda_{3\sigma}=0.31$, $\lambda_{5\sigma}=0.23$, $\lambda_{7\sigma}=0.2$ , $\lambda_{9\sigma}=0.18$ ,
 $\lambda_{11\sigma}=0.19$, $\lambda_{13\sigma}=0.01$\}. Using these fine tuned values of perturbation strengths, we get strong oscillations in the return probability (see Fig.\eqref{fig:vacstate}). Interestingly, even with such fine tuning, the spread complexity calculated from the naive Lanczos method does not show robust oscillations. However the same computed from FSA shows discernibly stronger oscillations over time. Evidently, these oscillations die out in the large timescale as iterative symmetry correction process via few body terms is only approximate in this case.

\section{Discussions and Conclusions}\label{sec6}
In the current work, we presented an exhaustive discussion of spread complexity for the kinematically constrained PXP systems. We focused largely on the protected sector of many-body scar states in the PXP spectrum that weakly violate the ETH. We computed quantum spread complexity starting from special initial states such as $|\mathbb{Z}_2\rangle$ and $|\mathbb{Z}_3\rangle$ that are known to host equidistant tower of scars.
We also discussed computation of complexity starting from the vacuum and accessing effective Floquet scar states nested atop. The challenge for computing spread complexity for PXP system is the non-closure of the symmetry algebra which strays away from $su(2)$, posing a difficulty in the Lanczos algorithm to stay in a finite Krylov subspace.
\medskip

We reiterated throughout this work, that in such cases where the scar subspace is not truly disjoint from bulk states, one must judiciously use Forward Scattering Approximation (FSA) as a better numerical alternative to designate associated Krylov subspace. Using FSA guaranteed the closure of the Lanczos algorithm, and hence a finite dimensional Krylov space. Our results cumulatively suggest (approximately) prohibiting the leakage of the wavefunction outside the Krylov subspace results in more dependable results for spread complexity. Evidently, all the FSA complexity values show well behaved bounded feature within our time-scale of observation as it can not access more and more higher order Krylov basis states, something prominently seen in bare Lanczos complexity. We also show that although the difference between these two approaches do not matter as such for optimal perturbation added to the Hamiltonian with $|\mathbb{Z}_2\rangle$ initial state, such instances are harder to find for other initial states. On top of that, the convergence of spread complexity calculated numerically via the naive method may converge much slower w.r.t number of Lanczos basis states in the bare case when compared to the optimal perturbed situation. See Appendix \eqref{apC} for some more comments on this. A careful analysis of the structure of the FSA errors and their scaling with system size provides important insight on the q-deformation of the emergent $SU(2)$ symmetry of the PXP model in the asymptotic limit.
\medskip

However there still remains a bunch of questions to address in this regime. Of course the most basic of them concerns the nature of the symmetry algebra for PXP systems. At the level of dynamics of eigenstates, one might conjecture that replacing the symmetry with $su(2)_q$ like quantum deformed structures might help. However, the PXP system cannot be exactly covered by this method for any value of $q$, and the better approximation regime is only reached when one adds optimal perturbations like in the $|\mathbb{Z}_2\rangle$ case. In that case $q$ is already almost unity, which defeats the whole purpose of this approximation. See Appendix \eqref{apB} for some more discussion on this. Moreover, for other initial states, say for $|\mathbb{Z}_3\rangle$, where the algebra is more broken than the $|\mathbb{Z}_2\rangle$ case, we tried to approximate Lanczos Coefficients using representation theory for further deformed algebras such as $su(2)_{p,q}$. This exercise turned out to be futile. Perhaps it is better to accept that the $su(2)$ algebra can be improved in more or less all cases in models like PXP, no matter how broken the representation is to begin with, by considering symmetry restoring terms of a cautiously defined broken representation.
\medskip

The present work is only a preliminary step towards understanding scar states from a spread complexity perspective and there are a number of further problems we can move on to at this point. Scar subspaces do exist in other interesting models, like the AKLT model \cite{MoudyagalyaRev} for instance, where the underlying dynamical symmetry is much more non-trivial than the PXP case. It would be interesting to see whether spread complexity is sensitive to subtleties associated to such kind of loosely embedded invariant subspaces. There are exciting new studies in semi-analytical perspectives as well, for example in \cite{Delacretaz2023} a $1+1$d $\phi^4$ QFT has been studied numerically, and several scar states have been found to coexist with thermalizing bulk states in the weak coupling regime. It will be interesting to see if a numerical computation of spread complexity gives us more insight into these states.

\section*{Acknowledgements} 
We thank Christopher J Turner for useful discussions. S. N. acknowledges support by the projects J1-2463 and P1-0044 program of the Slovenian Research Agency, EU via QuantERA grant T-NiSQ. B.M. has been funded by the European
Research Council (ERC) under the European Union’s Horizon
2020 Research and Innovation programme (Grant Agreement
No. 853368). A.B is supported by Mathematical Research Impact Centric Support Grant (MTR/2021/000490) by the Department of Science and Technology Science and Engineering Research Board (India) and Relevant Research Project grant (202011BRE03RP06633-BRNS) by the Board Of Research In Nuclear Sciences (BRNS), Department of Atomic Energy (DAE), India. AB thanks the participants of the workshop ``Quantum Information in QFT and AdS/CFT-III" organized at IIT Hyderabad and funded by SERB through a Seminar Symposia (SSY) grant (SSY/2022/000446) for useful discussions. The work of Aritra Banerjee (ArB) is supported by the Quantum Gravity Unit of the Okinawa Institute of Science and Technology Graduate University (OIST). ArB acknowledges the hospitality of Kyoto University and National Sun Yat Sen University (Kaohsiung) during various stages of this work. Some of the results were presented in a NSYSU-CTCP-NCTS joint seminar during ArB's visit to Taiwan.

\appendix

\section{FSA Errors}\label{apA}

As mentioned in \eqref{fsaerr}, the FSA error $\delta v_k$ for the k-th step is defined as follows 

\begin{equation}
    \delta v_k=||\mathcal{H}^{-}|v_k\rangle-\beta_k|v_{k-1}\rangle||
\end{equation}

where we have:
\begin{eqnarray}
    \ket{v_k}&=&\frac{1}{\beta_k}\mathcal{H}^+\ket{v_{k-1}}\nonumber\\
    \beta_k&=&\sqrt{||\mathcal{H}^+\ket{v_{k-1}}||}
\end{eqnarray}

and $\mathcal{H}^++\mathcal{H}^-=\mathcal{H}$ is the total Hamiltonian of the system. The precise form of $\mathcal{H}^{\pm}$ depends on from which initial state ($\ket{v_0}$) we want to start the FSA. First two FSA steps are error free both for $\ket{\mathbb{Z}_2}$ and $\ket{0}$. We now define average FSA error, for these two initial states, as  
\begin{equation}
\delta_{av}=\frac{\sum_{k=3}^{n^*+2}\delta v_k}{n^*}
\label{avg_lanc_error}
\end{equation}

where $n^*=\#$ FSA steps with nonzero errors. We have $n^*=L-3(L/2-2)$ for 
$\ket{\mathbb{Z}_2}(\ket{0})$. We note that, in addition to the first two steps, the last step is also error free for $\ket{\mathbb{Z}_2}$.
\medskip

Now we pursue a comparative study of the FSA errors for $\ket{\mathbb{Z}_2}$ and $\ket{0}$. To this end, we first  obtain analytical expressions of the FSA error (in terms of system size) in the 3rd step.  Let us do this first for the $\ket{\mathbb{Z}_2}$ initial state for which the PXP Hamiltonian is decomposed in the following manner

\begin{eqnarray}
    H&=&H^++H^-\nonumber\\
    &=&\sum_{i \ \text{odd}}\tilde{\sigma}^-_i+\sum_{i \ \text{even}}\tilde{\sigma}^+_i
\end{eqnarray} 

We start with $\ket{v_0}=\ket{\mathbb{Z}_2}=|101010\cdots\rangle$ and get 

\begin{eqnarray}
    \ket{v_1}&=&\frac{1}{\beta_1}H^+\ket{v_0}\nonumber\\
    &=&\sqrt{\frac{2}{L}}\left(\sum_{i \ \text{odd}}\tilde{\sigma}^-_i\ket{\mathbb{Z}_2}\right)
\end{eqnarray}
In the $\ket{\mathbb{Z}_2}$ state all odd sites are filled. So, the state $\ket{v_1}$ can be thought of as a translation invariant version of single hole (in the odd sites) states. Next, we get
\begin{eqnarray}
    \ket{v_2}&=&\frac{1}{\beta_2}H^+\ket{v_1}\nonumber\\
    &=&2\sqrt{\frac{2}{L(L-2)}}\sum_{\substack{i,j \ \text{odd}\\j>i}}\tilde{\sigma}^-_i\tilde{\sigma}^-_j\ket{\mathbb{Z}_2}
\end{eqnarray}

Now the second state $\ket{v_2}$ can be thought of as equal mixture of distinct 2-hole (in odd sites) states whose number is $L(L-2)/8$.
Following through, we find 
\begin{eqnarray}
    H^+\ket{v_2}&=&2\sqrt{\frac{2}{L(L-2)}}\Big(\sum_{\substack{i,j,k \ \text{odd}\\i\ne j\ne k}}\tilde{\sigma}^-_i\tilde{\sigma}^-_j\tilde{\sigma}^-_k\ket{\mathbb{Z}_2}\nonumber\\
    &&+\sum_{i=1}^{L/2}(T^2)^i\ket{010010101\cdots}\Big)
    \label{eq:Hpv2}
\end{eqnarray}
The first type of states are 3-hole states (in odd sites) and naively the number of such states generated under the action of $H^+$ on $\ket{v_2}$ is $\frac{L(L-2)}{8}(\frac{L}{2}-2)$. But, the number of distinct 3-hole states is $^{L/2}C_3$. So, the multiplicity factor of each 3-hole state in Eq.~\ref{eq:Hpv2} will be 3. $T$ is the translation operator by one lattice spacing. Thus, $\beta_3$ is given by
\begin{eqnarray}
    \beta_3&=&\sqrt{||H^+\ket{v_2}||}\nonumber\\
    &=&2\sqrt{\frac{2}{L(L-2)}}\sqrt{9\times ^{L/2}C_3+\frac{L}{2}}\nonumber\\
    &=&\sqrt{\frac{3(L-4)}{2}+\frac{4}{L-2}}
\end{eqnarray}
Finally we get the expression of $\ket{v_3}$ as $\ket{v_3}=\frac{1}{\beta_3}H^+\ket{v_2}$. To calculate the error in the 3rd step, one now need to calculate $H^-\ket{v_3}$ which is given by:
\begin{eqnarray}
    H^-\ket{v_3}&=&\frac{6\sqrt{2}}{\beta_3\sqrt{L(L-2)}}\Big(\frac{L-4}{2}\sum_{\substack{i,j \ \text{odd}\\j>i}}\tilde{\sigma}^-_i\tilde{\sigma}^-_j\ket{\mathbb{Z}_2}\nonumber\\
    &&+\frac{2\sqrt{2}}{\beta_3\sqrt{L(L-2)}}\sum_{i=1}^{L/2}(T^2)^i\ket{000010101\cdots}\Big)\nonumber\\
    \label{eq:Hmv3}
\end{eqnarray}
The states in the 2nd line of Eq.~\ref{eq:Hmv3} are specific kind of 2-hole states (in odd sites) which are also present in the group of 2-hole states in the first line. 
Thus, by carefully counting the newly generated states we obtain 
the FSA error in the 3rd step starting from $\ket{\mathbb{Z}_2}$ as 
\begin{eqnarray}
    \delta_3&=&||H^-\ket{v_3}-\beta_3\ket{v_2}||\nonumber\\
    &=&\left(\frac{3\sqrt{2}(L-4)}{\beta_3\sqrt{L(L-2)}}-\frac{2\sqrt{2}\beta_3}{\sqrt{L(L-2)}}\right)^2\times(\frac{L(L-2)}{8}-\frac{L}{2})\nonumber\\
    &&+\left(\frac{\sqrt{2}(3L-10)}{\beta_3\sqrt{L(L-2)}}-\frac{2\sqrt{2}\beta_3}{\sqrt{L(L-2)}}\right)^2\times\frac{L}{2}\nonumber\\
    &=&\frac{8(L-6)}{(L-2)(3L^2-18L+32)}
\end{eqnarray}

The expression of $\delta_3$ for the vacuum state ($\ket{0}$) can be found by a similar computation \cite{BhaskarPRB2020}, and it reads:
\begin{eqnarray}
    \delta_3&=&\frac{6(L-5)}{L^3-12L^2+47L-60}
\end{eqnarray}

\begin{figure*}
\includegraphics[width=\textwidth]{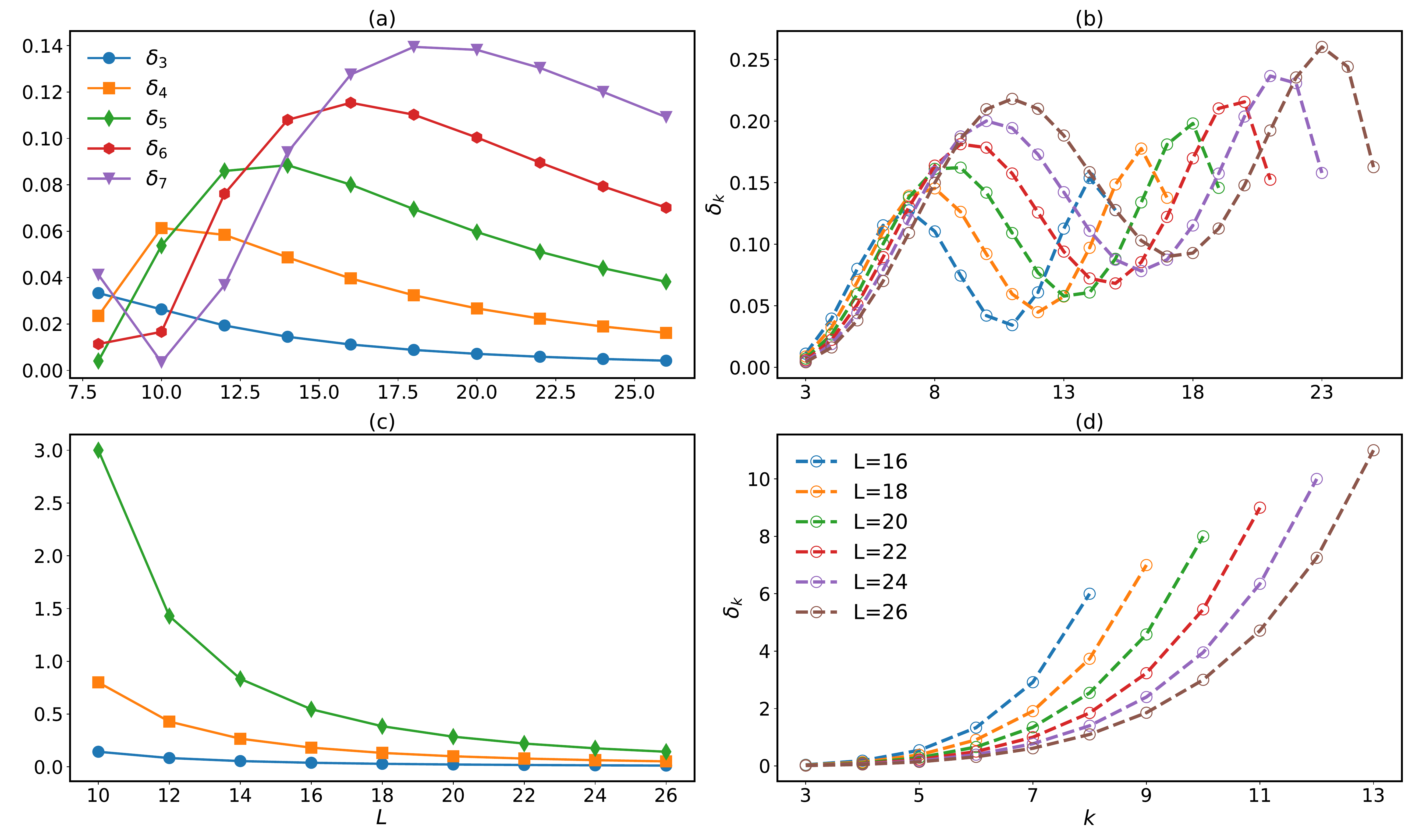}
\caption{FSA errors in various FSA steps for the $\mathbb{Z}_2$ state in the upper panels ((a),(b)) and for the vacuum state ($\ket{0}$) in the lower panels ((c),(d)). $\delta_k$ vs $L$ for different $k$ for $\mathbb{Z}_2$ (a) and $\ket{0}$ (c) and $\delta_k$ vs $k$ for different $L$ for $\mathbb{Z}_2$ (b) and $\ket{0}$ (d). }
\label{fig:fsa_errror}
\end{figure*}

One can see that though both the expressions scale as $1/L^2$ for large $L$, $\delta_3$ for $\ket{0}$ is larger compared to $\ket{\mathbb{Z}_2}$ at any finite size $L$. Another interesting feature to note is that $\delta_3=0$ at $L=6$ for $\ket{\mathbb{Z}_2}$. It then increases upto $L=8$ and then starts decreasing. In contrast $\delta_3$ for $\ket{0}$ is always monotonically decreasing ($L=6$ onwards ; the formula is not valid for $L<6$).
We numerically computed the behavior of $\delta_{n>3}$ for $\ket{\mathbb{Z}_2}$ ($\ket{0}$) and found the non-monotonic (monotonic) behavior with $L$ holds for $n>3$ as well (see Fig.\ref{fig:fsa_errror} (a),(c)). The non-monotonic (more precisely, oscillatory) nature of the FSA errors for the $\ket{\mathbb{Z}_2}$ case makes it difficult to foresee the behavior of $\delta_{av}$ (and hence the non-thermal dynamics from $\ket{\mathbb{Z}_2}$) in the thermodynamic limit though we find it to increase slowly with system size within our numerical capacity. The situation for $\ket{0}$ is also puzzling since $\delta_k$ for any $k$ is found to monotonically decrease with system size in spite of the fact that we see rapid thermalization from $\ket{0}$. Though we find that $\delta_k$, increases monotonically with $k$ for any system size (Fig.\ref{fig:fsa_errror}(d)) and it is $\delta_{L/2}$ (the error in the last step which interestingly found to have an integer value $L/2-2$ for even $L$) which makes $\delta_{av}$ to increase with $L$ (even) in this case. Note that the errors for $\ket{0}$ are more than one order of magnitude higher than the corresponding errors for the $\ket{\mathbb{Z}_2}$ case. Thus we find that the behavior of the FSA errors as a function of system size have a rich structure and provide important insight to understand the scaling of the dynamical behaviors with system size.

\begin{figure*}
\includegraphics[width=\textwidth]{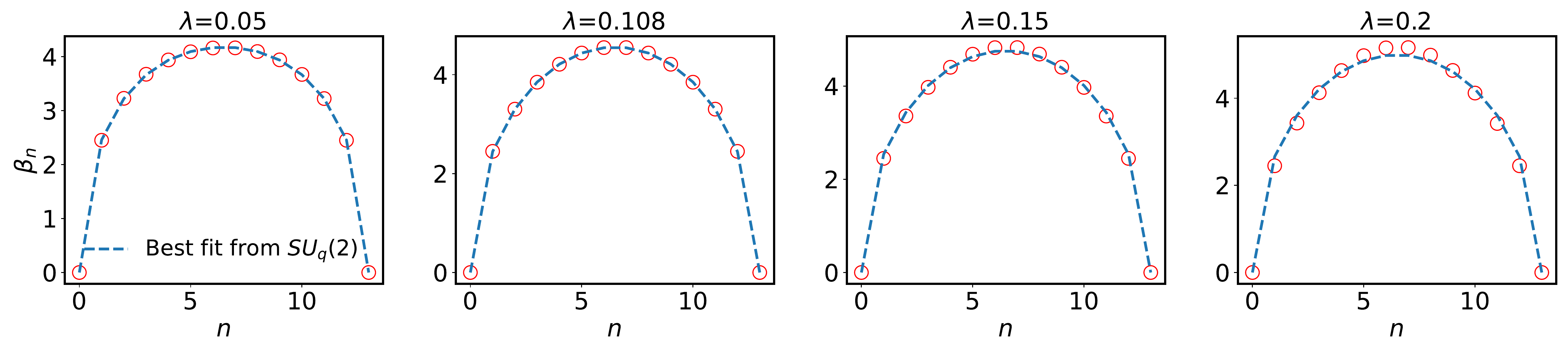}
\caption{Shows the $\b_n$ for FSA as a function of $n$ (i.e., iterations) with $|\Psi\rangle_{in}=|Z_2\rangle$ and $\mathcal{H} = \mathcal{H}_{PXP} + \lambda \mathcal{H}_{P, Z_2}$ for different perturbation. For $\lambda =0.05, 0.108$, the curve is consistent with a broken $su(2)$ symmetry which can still be captured via $su_{q}(2)$ algebra with $q=0.85, 0.96$ respectively. For $\lambda=0.108$, closeness of $q$ with identity indicates a restoration of the $su(2)$ symmetry. Note that $q$ eventually seems to move towards identity with increasing $L$. On the other hand, we see that for $\lambda \gtrsim 0.15$ or so, even the $su(2)_q$ symmetry is broken. We show here tow such values of $\lambda=0.2, 0.5$, both of which clearly show that even $su(2)_q$ cannot anymore capture brokenness of $su(2)$.}
\label{fig:qbroken}
\end{figure*}

\section{Comments on convergence of complexity}\label{apC}

\begin{figure}[ht!]
\includegraphics[width=\columnwidth]{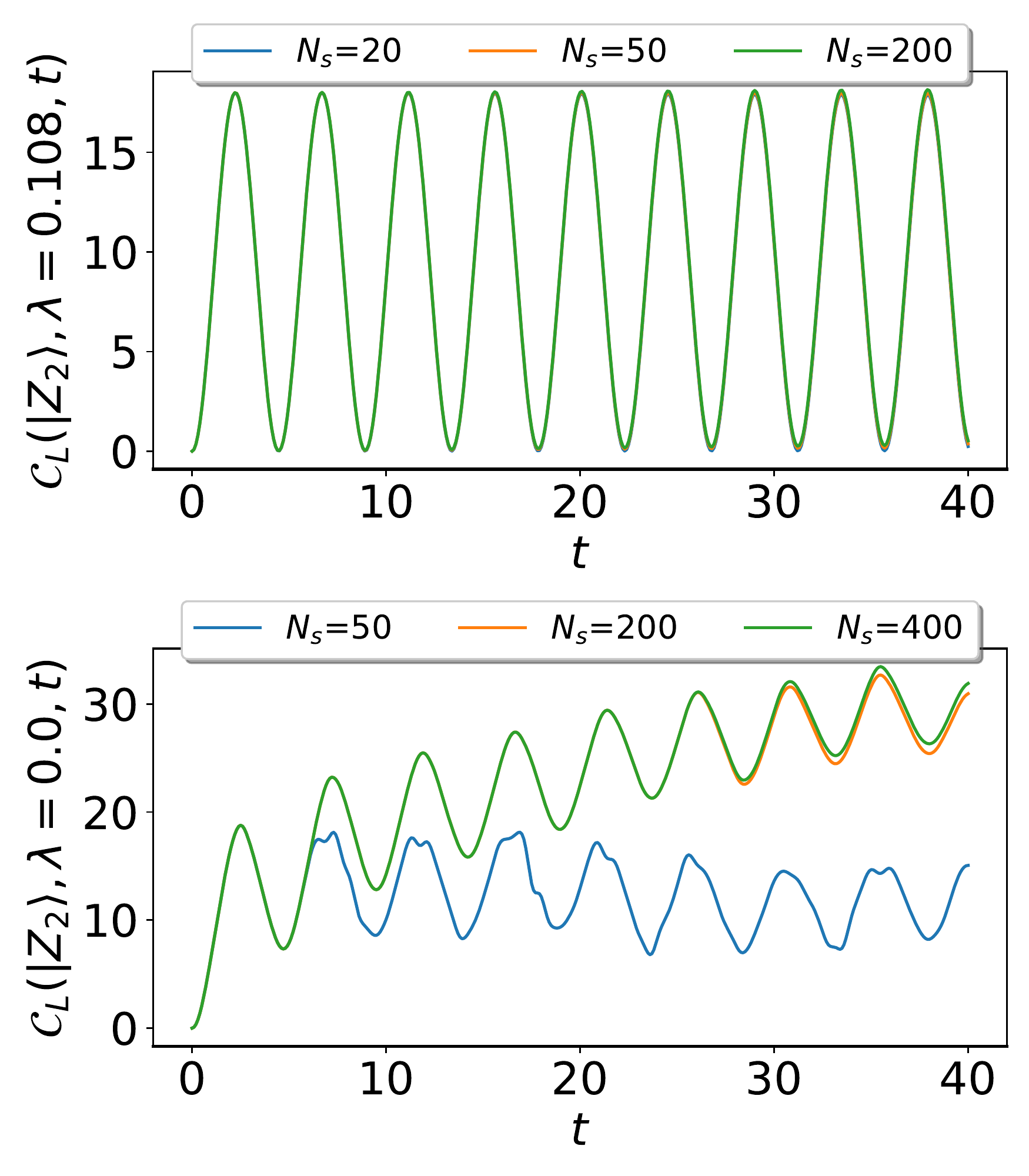}
\caption{Convergence of complexity w.r.t Lanczos steps plotted for $|\mathbb{Z}_{2}\rangle$ initial states with and without optimal perturbation strength. } \label{fig:Lanczos_convergence}
\end{figure}
As it is already demonstrated in Fig. \ref{fig:BCH_compare_b}(b) and Fig. \ref{fig:z3summary}(a), the naive Lanczos iterations do not terminate unlike the FSA. Therefore, it becomes rather imperative to check that the quantum complexity computed via Lanczos is indeed convergent w.r.t number of Lanczos basis states $N_{s}$. We discuss and show here the main results related to convergence of quantum complexity computed via Lanczos for $|\mathbb{Z}_{2}\rangle$ initial state. We checked that main conclusions we draw also hold for $|\mathbb{Z}_{3}\rangle$ and $|0\rangle$ states. To this end, we consider the perturbed PXP Hamiltonian $H=\mathcal{H}_{PXP} + \lambda \mathcal{H}'_{\mathbb{Z}_{2}}$. For $\lambda=0.108$ i.e., PXP system with optimal pertubation for revival from $|\mathbb{Z}_{2}\rangle$ states, we see that quantum complexity converges quite nicely for any number of Lanczos basis states $N_{s}>(L+1)$ as shown in upper panel of Fig. \ref{fig:Lanczos_convergence}.
On the other hand, for $\lambda=0$ which represents the unperturbed PXP model, one needs rather large number of Lanczos basis number $N_{s}$ such that $\mathcal{C}_{L}$ converges w.r.t $N_s$, as shown in lower panel of Fig. \ref{fig:Lanczos_convergence}. Furthermore, we observe that the required number of $N_s$ for a convergent value of $\mathcal{C}_{L}$ also depends on the time-scale and such conclusions hold for any other value of $\lambda$ which results in a non-optimal perturbation over PXP system in terms of revival from the $|\mathbb{Z}_{2}\rangle$ states. 
%Similar conclusions, as also pointed out earlier, holds for $|\mathbb{Z}_{3}\rangle$ and $|0\rangle$ initial state.
To summarize, for optimal perturbation that restores the broken $su(2)$ symmetry (to the best possible extent), $\mathcal{C}_{L}$ converges fast and it usually needs only $N_s \gtrsim N_{b}$ basis states where $N_{b}$ denotes the number of steps after which the corresponding FSA terminates. However, for any other value of perturbation strength, one needs to be careful while computing complexity via naive Lanczos method because one may require a much higher number of Lanczos steps i.e. basis states (also depending upon time) for the complexity to converge. Due to the guaranteed closure for the FSA method, the corresponding complexity computations do not suffer from such issues. 
%\newpage
 \section{Comments on $q$ deformation}\label{apB}

\begin{figure}[ht!]
\includegraphics[width=\columnwidth]{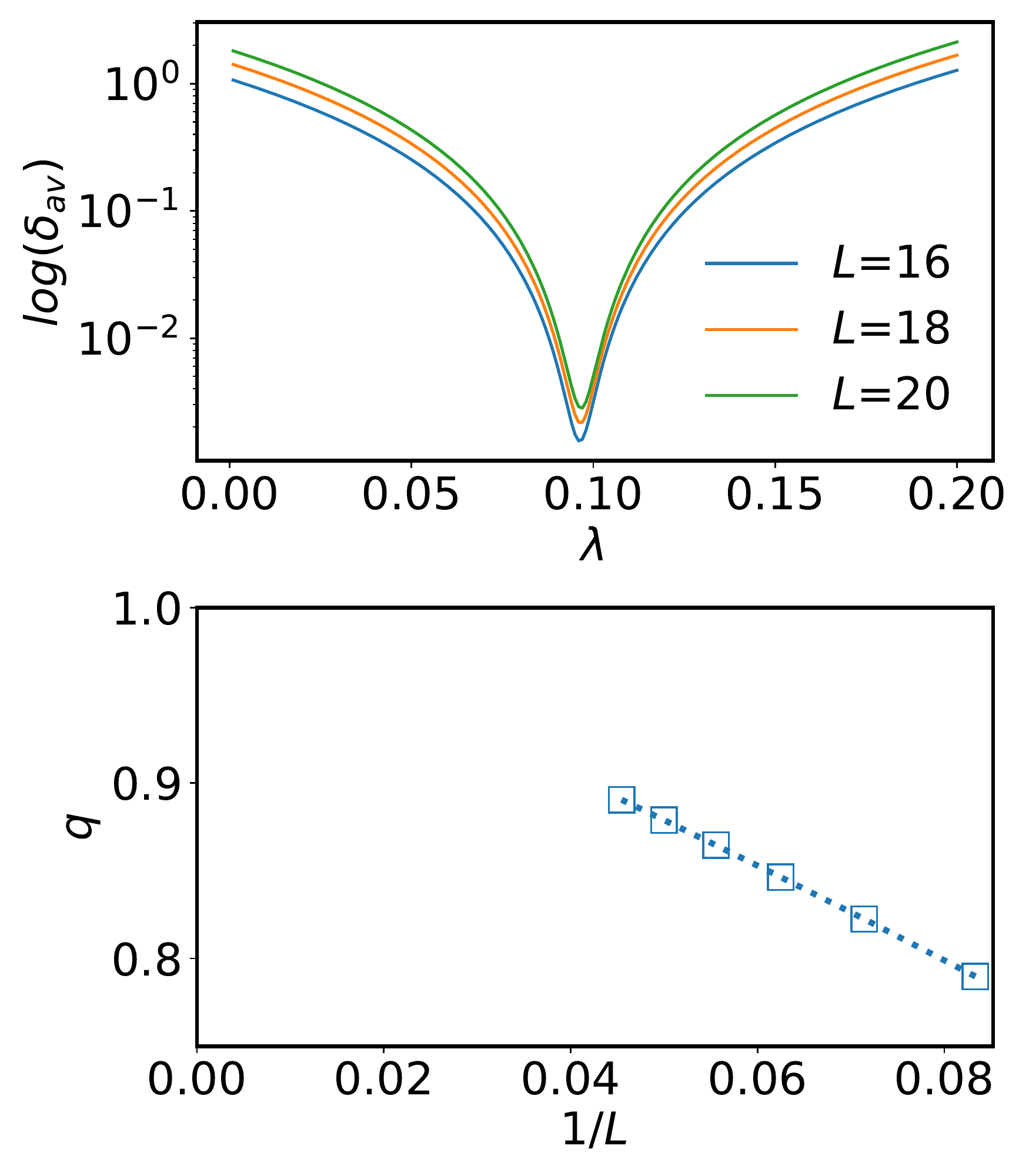}
\caption{Panel (a) shows the logarithm of the average Lanczos error $\delta_{av}$ as a function of the perturbation strength $\lambda$ for $L=16,18,20$ for $|\mathbb{Z}_2\rangle$ initial state and $\mathcal{H} = \mathcal{H}_{PXP} + \lambda \mathcal{H}_{P, Z_2}$. The fact that Lanczos error increases with $L$ clearly indicates that $su(2)$ symmetry cannot be restored in thermodynamic limit. This is important because if one analyses the $q$-numbers for such scenario, naive extrapolation w.r.t $L$ seems to suggest that $q \rightarrow 1$ for $L \rightarrow {\infty}$ which implies a restoration of perfect $su(2)$ symmetry. Therefore extrapolation w.r.t $L$ can be misleading and should be better avoided (as done in panel (b)) with the rather small $L$ we can compute for.}
\label{fig:qcompare}
\end{figure}
 
 In this subsection, we attempt to take an in-depth look at the idea of capturing the weakly broken $su(2)$ symmetry in PXP model via q-deformed $su(2)$ algebra (denoted by $su(2)_q$), as proposed in \cite{Bhattacharjee_2022}. In order to keep our discussion self-contained, let us first quickly review the basic concepts of $su(2)_q$ algebra. Usually, $q$ numbers are defined as a version of ordinary numbers parametrized by $q$ as 
\begin{eqnarray}
[x]_{q}=\frac{q^{x}-q^{-x}}{q-q^{-1}}
\label{qnumber_def}
\end{eqnarray}
It is clear from Eqn. \ref{qnumber_def} that $\lim_{q\rightarrow 1}[x]_{q}=x$. Now, $su(2)_q$ algebra is generated by the generators $J_{0}$ and $J_{\pm}$ satisfying the commutation relations
\begin{eqnarray}
[J_{0}, J_{\pm}] = \pm J_{\pm}, \hspace{4mm} [J_{+}, J_{-}]=[2J_{0}]_{q}
\end{eqnarray}
where $[J_{0}]_{q}$ is the q-deformed operator \cite{Bhattacharjee_2022}. Note that $su(2)_q$ is a \textit{non-linear} generalization of $su(2)$ and one recovers the $su(2)$ algebra for $q\to1$. As discussed in detail in \cite{Bhattacharjee_2022}, the Lanczos coefficients $\beta_{n}(q)$ for $su(2)_q$ are of the form
\begin{eqnarray}
\beta_{n}(q) = \alpha\sqrt{[n]_{q}[2j-n+1]_{q}}
\label{qLanc_coeffs}
\end{eqnarray}

where $\alpha$ is a constant and the natural basis of $su(2)_{q}$ algebra is given by $|j,n\rangle$ where $-j \le n \le j$. For the bare PXP model (i.e., in the absence of any perturbation), starting from the $|\mathbb{Z}_{2}\rangle$ state, one can try to fit the Lanczos coefficients with the expression \ref{qLanc_coeffs} and extract $q$ and $\alpha$. Both these parameters are $L$ (system size) dependent. Whereas $\alpha$ does not have any systematic behaviour w.r.t $L$, the behaviour of $q$ is plotted in Fig. \ref{fig:qcompare} lower panel.
Although the figure suggests that $q \rightarrow 1$ as $L \rightarrow \infty$ via \textit{naive extrapolation}, one needs to be more cautious before drawing such conclusion. Because, the conclusion of $q$ becoming 1 in the thermodynamic limit implies that $su(2)$ symmetry is exactly revived in the same. However, to this end, we check the average FSA error ( defined in the same manner as in Eqn. \ref{avg_lanc_error} but appropriately modified for FSA on PXP model with $|\mathbb{Z}_{2}\rangle$ initial state ) at different system sizes $L$ and the result is plotted in Fig. \ref{fig:qcompare} upper panel. Note that for our current discussion, we are only interested in $\lambda=0$. The aforementioned figure clearly shows (for any $\lambda$ however) that FSA error increases with system size L (albeit very slowly) and thus hinting that $su(2)$ symmetry cannot be revived in the thermodynamic limit as suggested by behaviour of $q$ w.r.t $L$. This apparent dichotomy, in our opinion, implies that for the system sizes $L$ we considered are yet to reach a proper `scaling regime' to draw conclusion from the functional dependence of $q$ on $L$ and thus naive extrapolation on the $q$ vs. $1/L$ curve may be misleading. 
\medskip

Continuing on the same formalism, it is imperative to see how effective this picture of $su(2)_q$ symmetry is in the case of perturbed PXP model having the form $\mathcal{H} = \mathcal{H}_{PXP} + \lambda \mathcal{H}'_{P, Z_2}$. As shown in Fig. \ref{fig:qbroken}, the approximate $su(2)_q$ symmetry indeed captures the behaviour of Lanczos coefficients for small perturbations such as $\lambda=0.05, 0.108$. In particular, for $\lambda=0.108$, we get $q=0.96$ which indicates the strongest revival of $su(2)$ symmetry in the presence of optimal perturbation as expected; whereas $q$ values are away from 1 in the case of $\lambda=0.05$. Nonetheless, as one keeps on increasing the strength of perturbation ($\lambda \gtrsim 0.15$), the whole picture of broken $su(2)$ symmetry being captured by $su(2)_q$ with some $q \ne 1$ breaks down; which sets a parametric limitation to the picture of capturing breakage of $su(2)$ symmetry in perturbed PXP systems.

\begin{figure*}[ht!]
\includegraphics[width=\textwidth]{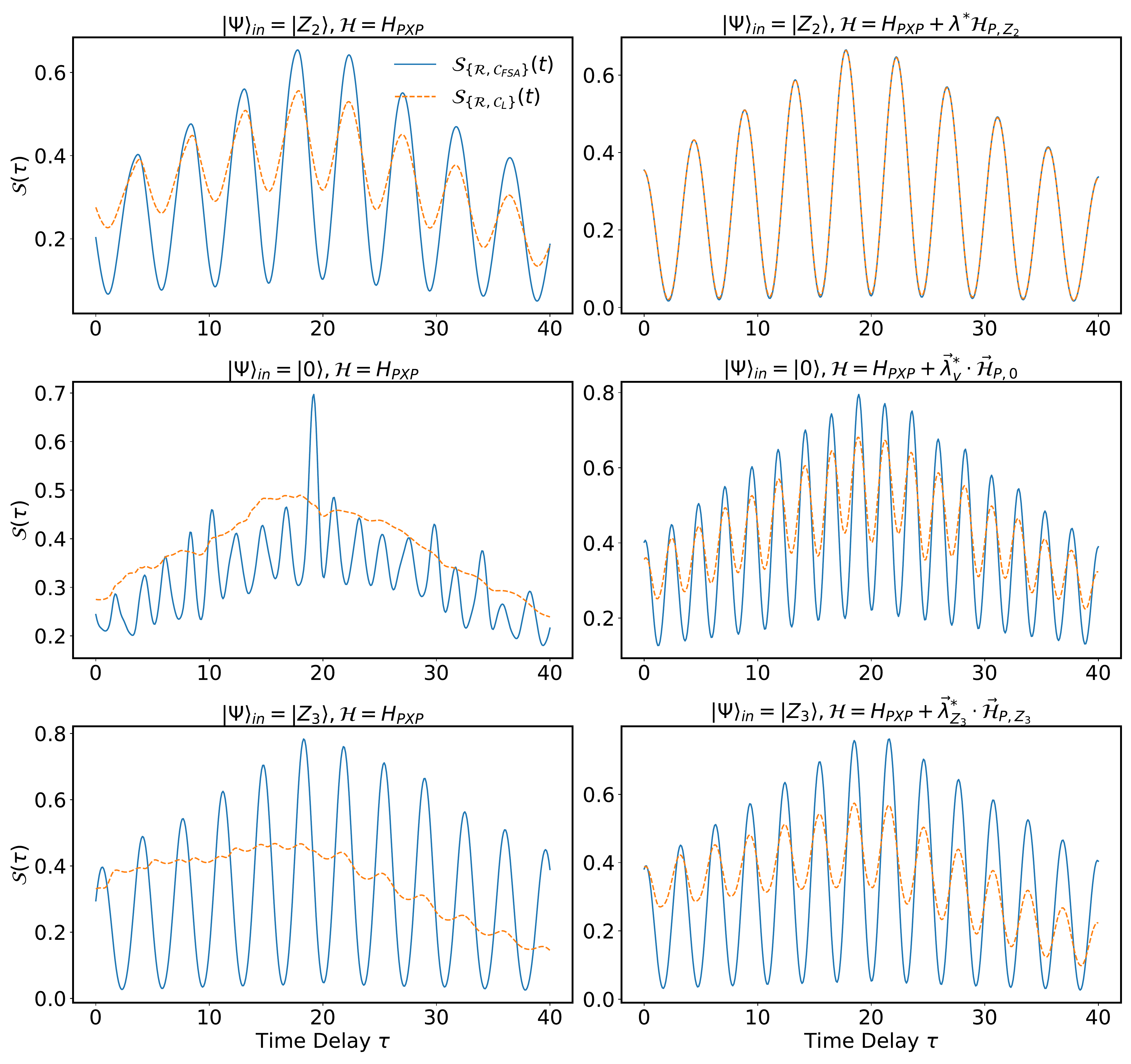}
\caption{Shows temporal behaviour of the cross correlation of $\mathcal{R}(t)$ w.r.t complexities $\mathcal{C}_{FSA}$ and $\mathcal{C}_{L}$ in case of both unperturbed and optimally casewise perturbed PXP model for $\mathbb{Z}_{2}\rangle$ (upper row), $|\mathbb{Z}_{3}\rangle$ (middle row) and vacuum $|0\rangle$ (lowest row) initial states.
As clearly seen, except for optimally perturbed PXP model with $|\mathbb{Z}_{2}\rangle$ initial state, in all cases $\mathcal{C}_{FSA}$ turns out to outperform $\mathcal{C}_{L}$ in terms of capturing similarity and dissimilarity w.r.t $\mathcal{R}(t)$, in the sense discussed in the text.}
\label{fig:Conv}
\end{figure*}

\section{Measuring cross-correlation}
We briefly comment on the cross-correlation between the temporal behaviour of return probability $\mathcal{R}(t)$ and the quantum complexity computed via both FSA and vanilla Lanczos methods in our work. Cross-correlations, denoted by $\mathcal{S}_{\{\mathcal{A},\mathcal{B}\}}$, between two sequences $\mathcal{A}$ and $\mathcal{B}$, essentially measures the `similarity' between the mentioned sequences. For our case, the sequences are essentially (discrete) time-series data and we follow the following definition of cross-correlation.

\begin{eqnarray}
\mathcal{S}_{\{\mathcal{A},\mathcal{B}\}}(\tau) = \sum_{n}\mathcal{A}_{n+\tau}\mathcal{B}_{\tau}
\end{eqnarray}
where $\tau$ denotes the lag or time delay. In Fig. \ref{fig:Conv}, we compare and contrast the cross-correlation of the return probability w.r.t quantum complexity $\mathcal{C}_{FSA}$ and $\mathcal{C}_{L}$ respectively extracted via FSA and Lanczos methods for different initial states and Hamiltonian containing optimal and non-optimal perturbation terms in regards to revival. As clearly seen in the figure, except for the case of optimally perturbed system for $|\mathbb{Z}_{2}\rangle$ revival, it is always $\mathcal{C}_{FSA}$ that outperforms $\mathcal{C}_{L}$ in terms of capturing its similarity and dissimilarity w.r.t return probability $\mathcal{R}$. This, in some sense, makes $\mathcal{C}_{FSA}$ the preferable candidate for capturing revivals in scarred systems.

\newpage
\bibliographystyle{apsrev4-1}

\bibliography{main}

\end{document}